\documentclass[showpacs,preprintnumbers,amsmath,amssymb,12pt,prb,floatfix]{revtex4}     
\usepackage{graphicx}
\usepackage{dcolumn}

\begin{document}

\title{A six-dimensional H$_2$--H$_2$ potential energy surface
for bound state spectroscopy}

\author{Robert J.\ Hinde}

\email{rhinde@utk.edu}

\affiliation{Department of Chemistry, University of Tennessee,
Knoxville, Tennessee 37996-1600}

\begin{abstract}

We present a six-dimensional potential energy surface for the
(H$_2$)$_2$ dimer based on coupled-cluster electronic structure
calculations employing large atom-centered Gaussian basis sets
and a small set of midbond functions at the dimer's center of
mass.  The surface is intended to describe accurately the bound
and quasibound states of the dimers (H$_2$)$_2$, (D$_2$)$_2$, and
H$_2$--D$_2$ that correlate with H$_2$ or D$_2$ monomers in the
rovibrational levels $(v, j) = (0, 0)$, (0, 2), (1, 0), and (1,
2).  We employ a close-coupled approach to compute the energies
of these bound and quasibound dimer states using our potential
energy surface, and compare the computed energies for infrared
and Raman transitions involving these states with experimentally
measured transition energies. We use four of the experimentally
measured dimer transition energies to make two empirical
adjustments to the ab initio potential energy surface; the
adjusted surface gives computed transition energies for 56
experimentally observed transitions that agree with experiment to
within 0.036~cm$^{-1}$.  For 29 of the 56 transitions, the
agreement between the computed and measured transition energies
is within the quoted experimental uncertainty. Finally, we use
our potential energy surface to predict the energies of another 34
not-yet-observed infrared and Raman transitions for the three
dimers.

\vskip\baselineskip

\noindent
Submitted to J.\ Chem.\ Phys.\ 22 Aug 2007; accepted
27 Nov 2007

\end{abstract}

\pacs{34.20.Cf, 31.25.-v, 36.40.-c}

\maketitle

\section{Introduction}

The (H$_2$)$_2$ dimer has long been viewed as a prototypical
bimolecular van der Waals dimer.  Because the (H$_2$)$_2$ dimer
is electronically simple, it has been the focus of a number of ab
initio studies;\cite{old1,old2,old3,old4,old5,old6,old7,old75,%
old8,old9,old10,old11,old12,jkj1,jkj2,old13,bmkp} however, because the
H$_2$--H$_2$ van der Waals interaction is quite weak,\cite{sk} ab
initio calculations with accuracy much higher than the oft-quoted
``chemical accuracy'' of 1~kcal/mol must be employed to provide
useful information about the H$_2$--H$_2$ potential energy
surface.  Two recent advances in ab initio methods have made it
possible to compute the H$_2$--H$_2$ interaction with the
required level of accuracy: (1) the development of hierarchical
sequences of one-electron Gaussian basis sets for approximating
molecular electronic wave functions,\cite{hbas} sequences which
systematically approach the complete one-electron basis set limit, and (2) the
development of efficient methods for accounting for electron
correlation effects in these wave functions by systematically
approaching the many-electron basis set limit.\cite{cc1,cc2}

Diep and Johnson\cite{jkj1,jkj2} took advantage of these two advances
in ab initio methods to compute an accurate four-dimensional
rigid rotor potential energy surface for the (H$_2$)$_2$ dimer;
the Diep--Johnson surface gives low temperature second virial
coefficients and integral elastic scattering cross sections in
reasonably good agreement with experiment.  However, this
potential energy surface does not depend explicitly on the
covalent bond lengths of the two H$_2$ monomers, and thus is only
able to describe the interaction between two H$_2$ molecules in
their $v=0$ vibrational ground states.

More recently, Boothroyd
{\it et al.\/}\cite{bmkp} have compiled a large database of
energies for the H$_4$ system, based largely on multireference
configuration interaction ab initio calculations, and have fit
a global six-dimensional H$_4$ potential energy surface to these
energies.  However, in this database, (H$_2$)$_2$
dimer configurations representative of the van der Waals well
are assigned energies that come not from ab initio
calculations, but rather from a empirically modified
rigid rotor potential energy surface.  Recent theoretical studies
of low-energy inelastic H$_2$--H$_2$ collisions that use this potential
energy surface\cite{clary,mate,lee} yield computed energy transfer
rate coefficients in rather poor agreement with experiment.

The (H$_2$)$_2$ dimer has also been the focus of several experimental
investigations, beginning with the pioneering work of Watanabe and
Welsh\cite{ww} that demonstrated the dimer's existence through
observation of its infrared (IR) absorption spectrum in the H$_2$ $v = 1
\leftarrow 0$ vibrational fundamental band.  Later experimental
studies\cite{mck1,mck2} recorded at high resolution the
IR absorption spectra of the (H$_2$)$_2$ dimer (and several
of its isotopomers) in the $v = 1 \leftarrow 0$ fundamental band and $v = 2
\leftarrow 0$ first overtone band of the corresponding monomers.
The high resolution IR absorption
spectra of (H$_2$)$_2$ in the H$_2$ fundamental and overtone regions,
and the analogous isotopomer spectra, provide information about the
vibrational dependence of the H$_2$--H$_2$ interaction.
Complementary studies\cite{mck3} of the far-IR absorption
spectrum of the dimer provide information about the anisotropy of
the potential energy surface in the region of the van der Waals
well.

Recently, the Raman spectrum of the (H$_2$)$_2$ dimer in the
H$_2$ fundamental region has also been observed.\cite{tej}  This
spectrum provides information about the vibrational dependence of
the H$_2$--H$_2$ interaction that is complementary to that
provided by the high-resolution IR studies.  Specifically, the
vibrationally excited state of (H$_2$)$_2$ that is probed by the
IR studies is one in which the vibrational excitation is
delocalized across the two H$_2$ monomers in an antisymmetric
fashion, while in the Raman studies, the excited (H$_2$)$_2$
state is one in which the vibrational excitation is delocalized
symmetrically across the two monomers.  A comparison of the IR
and Raman spectra thus provides insight into the coupling between
the two H$_2$ vibrational modes in the (H$_2$)$_2$ complex and
into the dependence of the H$_2$--H$_2$ potential energy surface
on the two monomers' bond lengths.

Equipped with this new information, we attempt here the
construction of a six-dimensional H$_2$--H$_2$ potential energy
surface that accurately describes the dimer's van der Waals well.
We begin by computing ab initio H$_2$--H$_2$ interaction energies
that are nearly converged with respect to both the one-electron
and many-electron basis sets, and then construct a smooth
potential energy surface from these computed interaction
energies.  We then make two small empirical adjustments to the
surface; these adjustments soften slightly the surface's
short-range repulsive wall, and increase slightly the strength of
the surface's anisotropic term that couples the rotational
degrees of freedom of the two monomers.  The empirically adjusted
surface gives IR and Raman transition energies for the ({\it
para\/}-H$_2$)$_2$, ({\it ortho\/}-D$_2$)$_2$, and {\it
para\/}-H$_2$--{\it ortho\/}-D$_2$ dimers in good agreement with
available experimental data.\cite{mck1,mck3,tej}

\section{Ab initio computations}

\subsection{Functional form of the H$_2$--H$_2$ interaction}

\label{sec:coeffs}

We consider a space-fixed coordinate system $(x, y, z)$ in which
one H$_2$ molecule (denoted molecule 1) has its center of mass at
the origin and the other H$_2$ molecule (denoted molecule 2) has
its center of mass on the positive $z$ axis.  The orientation of
molecule $i$ is specified by its spherical polar and azimuthal
angles $(\theta_i, \phi_i)$.  We let $R$ represent the distance
between the molecules' centers of mass, and let $r_i$ represent
the bond length of molecule $i$.  The H$_2$--H$_2$ potential
energy surface can then be expanded in terms of coupled spherical
harmonics:\cite{green}
   \begin{equation}
   \label{eq:fform}
   V =
   \sum_{l_1, l_2, L} A_{l_1, l_2, L}(R, r_1, r_2)
   G_{l_1, l_2, L}(\theta_1, \theta_2, \phi)
   \end{equation}
where $\phi = \phi_2 - \phi_1$, the summation indices $l_1$,
$l_2$, and $L$ are non-negative integers that must satisfy
   \begin{equation}
   l_1 + l_2 + L = {\rm even\ integer,}
   \end{equation}
and the homonuclear symmetry of the two H$_2$ monomers dictates
that $l_1$ and $l_2$ are also both even.  The angular functions
$G_{l_1, l_2, L}$ have the form
   \begin{equation}
   G_{l_1, l_2, L} = \sqrt{\frac{2 L + 1}{4 \pi}}
   \sum_m C(l_1, m, l_2, -m; L, 0) Y_{l_1,m}(\theta_1, \phi_1)
   Y_{l_2, -m}(\theta_1, \phi_2)
   \label{eq:geqn}
   \end{equation}
where $C$ is a Clebsch--Gordan coefficient and $Y_{l,m}$ is a
spherical harmonic normalized so that $Y_{l, m}(0, 0) =
\delta_{m,0} \sqrt{(2l + 1) / 4\pi}$.  (We use the Condon--Shortley
phase convention for $Y_{l,m}$.)  The appearance of
the Clebsch--Gordan coefficient $C$ in Eq.~(\ref{eq:geqn}) means
that $l_1$, $l_2$, and $L$ must satisfy the angular momentum
triangle rule.

The functions $G_{l_1, l_2, L}$ constitute a complete, orthogonal
basis set for functions of the three angular
coordinates $(\theta_1, \theta_2, \phi)$.  For fixed $R$, $r_1$,
and $r_2$, the coefficient $A_{l_1, l_2, L}(R, r_1, r_2)$
can therefore be computed as
   \begin{equation}
   \label{eq:project}
   A_{l_1, l_2, L}(R, r_1, r_2) = \frac{1}{2L+1}
   \int \int G_{l_1, l_2, L}(\theta_1,
   \theta_2, \phi) V(R, r_1, r_2, \theta_1, \theta_2, \phi)
   \, {\rm d}S_1 \, {\rm d}S_2 
   \end{equation}
where ${\rm d}S_i = \sin \theta_i \, {\rm d}\theta_i \, {\rm d}\phi_i$.

Earlier studies of the four-dimensional rigid-rotor
H$_2$--H$_2$ potential energy surface\cite{sk,jkj1,jkj2} show that the surface
is dominated by four terms, with $(l_1, l_2, L) = (0, 0,
0)$, $(0, 2, 2)$, $(2, 0, 2)$, and $(2, 2, 4)$.  In this
work, we use numerical quadrature to compute the right-hand
side of Eq.~(\ref{eq:project}) for these four $(l_1, l_2,
L)$ triples.  Specifically, at fixed values of $R$, $r_1$, and
$r_2$, we use the 18-point spherical
quadrature rule numbered 25.4.64 in Ref.~\onlinecite{as}
to evaluate the integrals over both ${\rm d}S_1$ and ${\rm d}S_2$
in Eq.~(\ref{eq:project}).
This requires us to compute the H$_2$--H$_2$ interaction energy
$V(R, r_1, r_2, \theta_1, \theta_2, \phi)$, using ab initio quantum
chemical methods that we describe in the next subsection,
at 12 sets of angles $(\theta_1, \theta_2, \phi)$ when $r_1 = r_2$
and at 19 sets of angles when $r_1 \ne r_2$.  Symmetry relationships
allow the rest of the $18^2 = 324$ interaction energies at fixed
$(R, r_1, r_2)$ to be determined from these ab initio calculations.

The accuracy of the $A_{l_1, l_2, L}$ coefficients computed
in this fashion is limited by the fact that the quadrature
rule we use fails to reproduce the orthogonality conditions
   \begin{equation}
   \label{eq:orthonorm}
   \int \int G_{l_1, l_2, L}(\theta_1,
   \theta_2, \phi) G_{l_1^\prime, l_2^\prime, L^\prime}(\theta_1,
   \theta_2, \phi)  \, {\rm d}S_1 \, {\rm d}S_2 
   = \delta_{l_1, l_1^\prime} \delta_{l_2, l_2^\prime}
   \delta_{L, L^\prime} (2L + 1)
   \end{equation}
when $l_1 + l_1^\prime \ge 6$ or $l_2 + l_2^\prime \ge 6$.  This
means that the value of $A_{0,0,0}$ obtained via quadrature also
includes some contamination from $A_{6,0,6}$ and $A_{0,6,6}$ (if
these coefficients are nonzero in the ab initio potential energy
surface), while $A_{2,2,4}$ is contaminated by (among other
terms) $A_{2,4,6}$ and
$A_{4,2,6}$, which describe the long-range electrostatic
quadrupole--hexadecapole (QH) interaction between the two H$_2$
molecules.

To assess the magnitude of these
erroneous contributions to the four $A_{l_1, l_2, L}$ coefficients
of interest, we used the more accurate 24-point spherical quadrature
rule of Ref.~\onlinecite{as} to calculate the coefficients at
$(R, r_1, r_2) = (4.5~a_0, 1.4~a_0, 1.7~a_0)$, a repulsive
(H$_2$)$_2$ configuration where we expect
the angular anisotropy of the potential energy surface to be relatively
high, and where this contamination should thus be relatively
severe.  Table~\ref{tab:one} compares the coefficients obtained
using the two quadrature rules (based on ab initio interaction
energies computed using the protocol outlined in Sec.~\ref{sec:pert});
the errors introduced at this $(R, r_1, r_2)$ configuration
by using 18-point quadrature
appear to be quite small for the four terms that we include in
our final potential energy surface.
This table also gives the values for two additional coefficients in the coupled
spherical harmonic expansion, $A_{2,2,0}$ and
$A_{2,2,2}$, at this (H$_2$)$_2$ configuration, and shows that
they are one to two orders of magnitude smaller than any of
the four terms we retain in Eq.~(\ref{eq:fform}).  This is in
accord with previous studies\cite{sk,jkj1,jkj2} of the four-dimensional rigid-rotor
H$_2$--H$_2$ potential energy surface.

\subsection{CCSD(T) ab initio calculations}

\label{sec:pert}

We use Gaussian 03\cite{g03} to compute the H$_2$--H$_2$
interaction energy, employing a coupled-cluster\cite{cc1,cc2}
treatment of electron correlation that includes single and double
excitations and a perturbative treatment of triple
excitations,\cite{pert} abbreviated CCSD(T).  The CCSD(T)
calculations are based on a restricted Hartree--Fock reference
wave function; we have verified that such a reference does not
exhibit a restricted $\rightarrow$ unrestricted instability for
the H$_2$ bond lengths considered here.  We use the
aug-cc-pVQZ basis set\cite{hbas,hbas2} for the four hydrogen
atoms, supplement this atom-centered basis set with a set of
(3s3p2d) bond functions positioned at the dimer's center of mass,
and employ the standard counterpoise correction.\cite{bbcp} The
bond function exponents are taken from Ref.~\onlinecite{tao}.

We carry out these calculations at $r_1$ and $r_2$ values of
$1.1$, $1.4$, and $1.7~a_0$, and at 19 $R$ values ranging from $R
= 4.25~a_0$ to $12.0~a_0$, for a total of 1653 unique $(R, r_1,
r_2, \theta_1, \theta_2, \phi)$ (H$_2$)$_2$ configurations.  We
turn off automatic checking of the one-electron overlap matrix
for near linear dependence and retain all 206 one-electron basis
functions at every configuration; this eliminates possible
discontinuities in the potential energy surface that could arise
when some of these functions are dropped from the one-electron
basis set. The Gaussian 03 H$_2$--H$_2$ total CCSD(T) energies
for these configurations are available from the EPAPS
depository.\cite{epaps}  We have checked a small subset of
these energies against calculations using the Dalton
ab initio code;\cite{dal} the dimer total energies computed
using the two codes agree to within $2 \times
10^{-8}$~hartrees or better.

To assess the error introduced by truncating the one-electron
basis set at the aug-cc-pVQZ + (3s3p2d) level, we performed
some calculations at selected configurations using a smaller
aug-cc-pVTZ atom-centered basis set and the same (3s3p2d) bond
function set.  The coefficients $A_{l_1, l_2, L}$ obtained
from these two sets of ab initio interaction energies are
listed in Table~\ref{tab:two}.  The two sets of coefficients
generally differ by no more than 1\% to 2\%, suggesting that
the aug-cc-pVQZ + (3s3p2d) basis set is nearly saturated.
Truncating the one-electron basis set seems to have the largest
effect on the isotropic coefficient $A_{0,0,0}$ computed at
small values of $R$, where the potential energy surface is
strongly repulsive.

\subsection{CCSDT ab initio calculations}

Our earlier study of the vibrational dependence of the H$_2$--H$_2$
interaction\cite{fbs} indicates that incompleteness in the many-electron
basis set could materially affect the shape of the potential energy
surface in the van der Waals well.  Similar
effects have been observed in other weakly-bound dimers of two-valence-electron
systems.\cite{hehe,mghe}  To reduce the error associated with truncation
of the many-electron basis set at the CCSD(T) level of theory, we
employ a coupled-cluster treatment that includes a fully iterative
treatment of single, double, and triple excitations,\cite{tripA,tripB}
abbreviated CCSDT, to compute the H$_2$--H$_2$ interaction energy
at selected high-symmetry geometries (those in which $\theta_1$,
$\theta_2$, and $\phi$ take values of 0 or $\pi/2$).  These
calculations are performed using the tensor contraction
engine\cite{tce} incorporated into version 4.7 of the electronic
structure code NWChem.\cite{nwc,nwc2}

Unfortunately, the CCSDT calculations
are prohibitively expensive if we employ the aug-cc-pVQZ + (3s3p2d)
one-electron basis set used in the CCSD(T) calculations.  We therefore
perform the CCSDT calculations using a smaller one-electron
basis set consisting of only atom-centered aug-cc-pVTZ functions.
We also use NWChem to perform CCSD(T) calculations at these
high-symmetry geometries using the atom-centered aug-cc-pVTZ
basis set.  We then take the
difference between the CCSDT and CCSD(T) counterpoise-corrected
interaction energies as an additive correction to the
aug-cc-pVQZ + (3s3p2d) CCSD(T) potential energy surface.
For the sake of brevity, we will call this the ``full-triples''
correction.  We found that to insure convergence of the CCSDT
iterations at some geometries, it was necessary to increase
the cutoff for computational linear dependence in the one-electron
basis set to $10^{-6}$.  For consistency, we therefore used this
cutoff in all of the CCSDT and CCSD(T) calculations performed
with NWChem.

Because we compute the full-triples correction at a small number of
H$_2$--H$_2$ orientations $(\theta_1, \theta_2, \phi)$, we cannot
use the quadrature scheme described in the the previous subsection
to extract corresponding full-triples corrections to the $A_{l_1, l_2, L}$
coefficients computed at the CCSD(T) aug-cc-pVQZ + (3s3p2d) level
of theory.  Instead, we use least-squares techniques to fit
the full-triples correction to the function
   \begin{eqnarray}
   \nonumber
   \lefteqn{
   \Delta A_{0, 0, 0}(R, r_1, r_2)
   G_{0, 0, 0}(\theta_1, \theta_2, \phi) +
   \Delta A_{0, 2, 2}(R, r_1, r_2)
   G_{0, 2, 2}(\theta_1, \theta_2, \phi)} \\
   & & {} + \Delta A_{2, 0, 2}(R, r_1, r_2)
   G_{2, 0, 2}(\theta_1, \theta_2, \phi) +
   \Delta A_{2, 2, 4}(R, r_1, r_2)
   G_{2, 2, 4}(\theta_1, \theta_2, \phi) \ .
   \label{eq:ftrip}
   \end{eqnarray}
We then add the corrections $\Delta A_{l_1, l_2, L}$ to
the corresponding coefficients $A_{l_1, l_2, L}$ obtained
from four-dimensional quadrature over the CCSD(T) aug-cc-pVQZ
+ (3s3p2d) interaction energies.
The CCSDT and CCSD(T) energies
used to compute the full-triples correction are available
through EPAPS.\cite{epaps}
For the sake of brevity,
we henceforth use the term ``coefficients'' to mean the sum
of the CCSD(T) coefficients and the full-triples corrections.

\subsection{Construction of a smooth potential energy surface}

\label{sec:smooth}

We now construct a smooth potential energy surface from the ab
initio coefficients $A_{l_1, l_2, L}(R, r_1, r_2)$.  For each
pair of H$_2$ bond lengths $(r_1, r_2)$, we create four cubic
splines, one for each of the coefficients $A_{l_1, l_2, L}$, that
interpolate the 19 coefficient values between $R = 4.25~a_0$ and
$R = 12.0~a_0$.  We extrapolate the splines to $R$ values below
$4.25~a_0$ and above $12.0~a_0$ using functions described in
the next two paragraphs.  At $R = 4.25~a_0$, the slope of each cubic
spline is constrained to match the slope of the corresponding
small-$R$ extrapolating function.  

We extend each cubic spline to $R$ values below
$4.25~a_0$ using a simple two-parameter exponential extrapolation
of the form $U \exp(-c R)$ that fits the coefficients obtained at
$R = 4.25~a_0$ and $4.5~a_0$.  We should stress that this extrapolation
is {\it not\/} expected to give highly accurate interaction
energies for small $R$; we use it simply to define the
slope for the cubic spline at $R = 4.25~a_0$.  The dimer
bound state wave functions we compute using our potential
energy surface are not sensitive to the highly repulsive
small-$R$ region of the potential energy surface.

Beyond $R = 12.0~a_0$, we extrapolate each spline using an
inverse-power expansion of the form $\sum_n C_n / R^n$, including
terms with $n = 5$ and 6 in the extrapolations for
$A_{2,2,4}$, terms with $n = 6$, 8, and 10 for $A_{0,0,0}$,
and terms with $n = 6$ and 8 in the extrapolations
for $A_{0,2,2}$ and $A_{2,0,2}$.  All $C_n$ coefficients
are determined as functions of $r_1$ and $r_2$.  The $C_5$ coefficient for
$A_{2,2,4}$ is computed from the H$_2$ quadrupole moments
listed in Ref.~\onlinecite{quad}.  The $C_6$ coefficients are obtained from
the isotropic and anisotropic $R^{-6}$ dispersion energy
coefficients given in Ref.~\onlinecite{disp} and the expressions
given in Ref.~\onlinecite{stone}.  The $C_8$ and $C_{10}$
dispersion energy coefficients are obtained from Ref.~\onlinecite{tim}.

To reduce the discontinuities in the higher-order derivatives
of the coefficients at $R = 12.0~a_0$, where the cubic spline
meets the long-range inverse-power extrapolating function, we
use the long-range function to compute values of the $A_{l_1, l_2, L}$
coefficents at six evenly spaced ``phantom`` points ranging from
$R = 13.0~a_0$ to $R = 18.0~a_0$, and force the spline to
intercept these phantom points as well as the points computed
at the 19 $R$ values cited above.  At $R = 18.0~a_0$, we also
constrain the slope of the spline to match that of the
inverse-power expansion.  However, we only use the spline
to evaluate the coefficients between $R = 4.25~a_0$ and
$R = 12.0~a_0$; beyond $R = 12.0~a_0$, we use the inverse-power
expansion to compute the coefficients $A_{l_1,l_2,L}$.

Using these extrapolated cubic splines, we can compute the coefficients
$A_{l_1, l_2, L}(R, r_1, r_2)$ at any $R$ for the discrete pairs
of H$_2$ bond lengths $(r_1, r_2)$ at which we performed the ab
initio calculations described above.  As the last step in
defining a smooth potential energy surface, we fit these
interpolated (or extrapolated) coefficients to the expression
   \begin{equation}
   \label{eq:r1r2}
   \sum_{k=0}^2 \sum_{n=0}^2 c_{k,n}
   (r_1 - r_{\rm{eq}})^k (r_2 - r_{\rm{eq}})^n
   \end{equation}
where $r_{\rm eq} = 1.4~a_0$.

\begin{figure}
\includegraphics{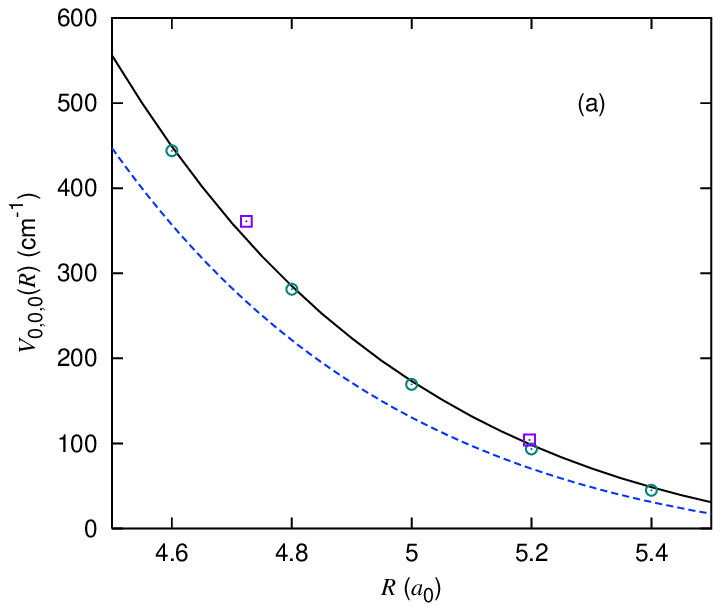}
\includegraphics{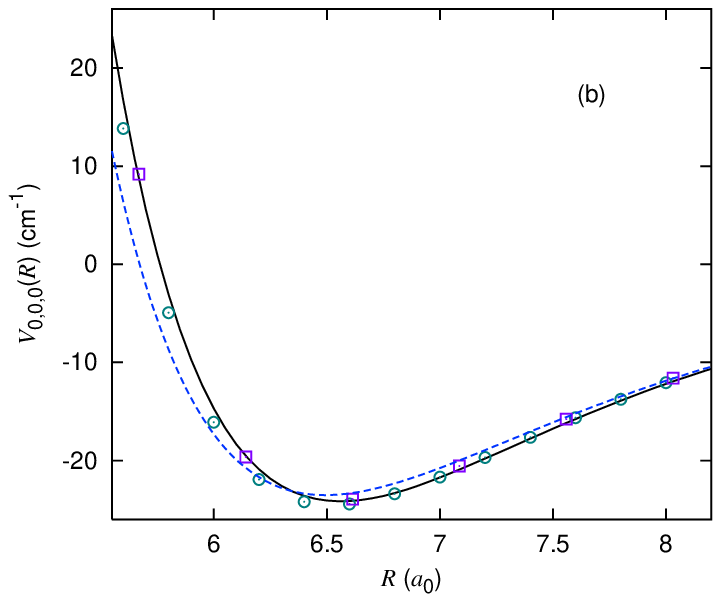}
\caption{(Color online.)
Comparison of the vibrationally-averaged isotropic H$_2$--H$_2$
potential energy curvce $A_{0,0,0}(R)$ obtained in this work (solid line) with the
isotropic rigid-rotor potential energy curves obtained
by other researchers.  Boxes represent the
extrapolated CCSD(T) potential energy surface of Refs.~\onlinecite{jkj1} and
\onlinecite{jkj2};
circles represent the adjusted ab initio potential of Ref.~\onlinecite{sk};
the dashed line represents the empirical potential of Ref.~\onlinecite{buck}.
The solid line shown here is computed from the unmodified $(s, q) = (0, 0)$
potential energy surface.
Panel (a) shows the repulsive wall at small $R$
values; panel (b) shows the van der Waals well.  }
\label{fig:repul}
\end{figure}

Figure~\ref{fig:repul} shows how the isotropic coefficient
$A_{0,0,0}(R, r_1, r_2)$, vibrationally averaged over the ground
state vibrational wave functions of the two H$_2$ monomers,
depends on $R$ both in the small-$R$, repulsive region of the
potential energy surface and in the shallow H$_2$--H$_2$ well. We
compare the vibrationally-averaged $A_{0,0,0}$ coefficient
computed in this work with a modified ab initio potential energy
surface\cite{sk} that gives accurate predictions for the
low-temperature second virial coefficient of H$_2$ gas, with the
extrapolated CCSD(T) ab initio calculations of Diep and
Johnson,\cite{jkj1,jkj2} and with an empirical isotropic potential
energy
curve\cite{buck} obtained from an analysis of the total
scattering cross section of moderate energy H$_2$--D$_2$
collisions.

In the shallow well, our vibrationally averaged $A_{0,0,0}$
coefficient agrees fairly well with the extrapolated CCSD(T)
results,\cite{jkj1,jkj2} which were computed within the rigid-rotor
approximation using the $v = 0$
vibrationally averaged bond length for both H$_2$ monomers.
The repulsive wall of our isotropic potential energy curve is
slightly softer than that of the extrapolated CCSD(T) curve;
our repulsive wall closely tracks the shape of the modified ab initio potential
energy surface\cite{sk} that gives accurate second virial
coefficients, except that our repulsive wall is shifted to
slightly larger $R$ values.
It is interesting to note that in the
small-$R$ repulsive region, the empirical isotropic potential
energy
curve\cite{buck} derived from scattering data is considerably
softer than any of the three curves derived from ab initio
computations.

\section{Computation of dimer bound state energies}

We assess the quality of our potential energy surface by using
it to compute the energies of several bound (and long-lived
quasibound) states of the (H$_2$)$_2$, H$_2$--D$_2$, and (D$_2$)$_2$
dimers.  In this section, we summarize the methods used to compute
these energies; the energies themselves
are presented in later sections.  We employ a standard close-coupled
approach\cite{green} in which the nine-dimensional dimer wave function
is written as
   \begin{equation}
   \label{eq:ccdimer}
   \Psi({\bf R}, {\bf r}_1, {\bf r}_2) =
   R^{-1} \sum_{\lambda}
   F_{\lambda}(R) 
   I_{J,M,\gamma}(\hat{\bf R}, \hat{\bf r}_1, \hat{\bf r}_2)
   \phi_{v_1,j_1}(r_1) \phi_{v_2,j_2}(r_2) \ .
   \end{equation}
Here ${\bf r}_i$ is the vector separating the two nuclei of H$_2$
molecule $i$, ${\bf R}$ is the vector separating the two
molecules' centers of mass, and $\hat{\bf r}_i$ and $\hat{\bf R}$
are the corresponding unit vectors.  The quantum numbers $J$ and
$M$ represent respectively the total angular momentum of the
dimer (excluding nuclear spin angular momentum)
and its projection on a space-fixed $z$ axis.  The
angular basis functions $I_{J,M,\gamma}$, which are defined as
   \begin{eqnarray}
   \nonumber
   \lefteqn{
   I_{J,M,\gamma}(\hat{\bf R}, \hat{\bf r}_1, \hat{\bf r}_2) =
   \sum_{m_1, m_2, M_{12}, N}
   C(j_1, m_1, j_2, m_2; J_{12}, M_{12})
   } \\
   & & {} \times  C(J_{12}, M_{12}, L, N; J, M)
   Y_{j_1,m_1}(\hat{\bf r}_1) Y_{j_2,m_2}(\hat{\bf r}_2)
   Y_{L,N}(\hat{\bf R}) \ ,
   \label{eq:angbasis}
   \end{eqnarray}
couple the rotational angular momenta
$(j_1, j_2)$ of the two H$_2$ molecules with the orbital angular
momentum $L$ of the dimer to create functions of definite $J$ and
$M$; we use $\gamma$ to represent the collection of angular
momentum quantum numbers $(j_1, j_2, J_{12}, L)$, where $J_{12}$
is the quantum number corresponding to the (vector) sum of the
rotational angular momenta of the two H$_2$ molecules. The
summation index $\lambda$ represents a collection of eight
quantum numbers:  the four quantum numbers listed in $\gamma$,
the total angular momentum quantum numbers $J$ and $M$, and the
vibrational quantum numbers $v_1$ and $v_2$ of the two monomers. 
The functions $\phi_{v,j}(r)$ are H$_2$ monomer radial wave functions,
defined so that
   \begin{equation}
   \int_0^\infty \phi^*_{v,j}(r) \phi_{v^\prime,j^\prime}(r) \, {\rm d}r =
   \delta_{v,v^\prime} \delta_{j,j^\prime}
   \end{equation}
and obtained from a Numerov--Cooley\cite{nc} analysis of the Ko\l
os--Wolniewicz\cite{kw} H$_2$ potential energy curve.

The dimer radial functions $F_\lambda(R)$ are solutions to a set
of coupled second-order differential equations; the $R$-dependent
terms that couple the radial functions $F_\lambda(R)$ and
$F_{\lambda^\prime}(R)$ are obtained by integrating the
six-dimensional H$_2$--H$_2$ potential energy surface over the
eight coordinates $({\bf r}_1, {\bf r}_2, \hat{\bf R})$, and are
defined by replacing the rigid-rotor potential coefficients
$A_{l_1,l_2,L}(R)$ in Eq.~(9) of Ref.~\onlinecite{green} with the
corresponding vibrationally averaged coefficients
   \begin{eqnarray}
   \nonumber
   \lefteqn{
   \langle v_1, j_1; v_2, j_2 | A_{l_1,l_2,L}(R, r_1, r_2) | v_1^\prime, j_1^\prime;
   v_2^\prime, j_2^\prime \rangle =} \\
   & & {} \int_0^\infty
   \int_0^\infty
   \phi^*_{v_1,j_1}(r_1) \phi^*_{v_2,j_2}(r_2) A_{l_1,l_2,L}(R, r_1, r_2)
   \phi_{v_1,j_1}(r_1) \phi_{v_2,j_2}(r_2) \, {\rm d}r_1 \, {\rm d}r_2 \ .
   \end{eqnarray}
We use the ABVN program\cite{abvn} to evaluate the angular momentum
coupling coefficients that appear in Eq.~(9) of
Ref.~\onlinecite{green}.

We convert the set of coupled second-order differential equations
to a matrix eigenproblem by discretizing the equations on a grid
in $R$, ranging from $R_{\text{min}} = 3.0~a_0$ to
$R_{\text{max}} = 28.0~a_0$ in steps of $0.1~a_0$, and replacing
the dimer radial kinetic energy operator with a five-point
central difference approximation evaluated on the grid. 
(Convergence tests show that using a smaller step size or a
larger value of $R_{\text{max}}$ does not change significantly
the energies of the
dimer states considered here.) We then solve the matrix
eigenproblem using the ARPACK code\cite{arpack} driven by the
SYMMLQ linear algebra routine.\cite{symmlq}  We truncate the wave
function expansion given in Eq.~(\ref{eq:ccdimer}) by limiting
$j_1$ and $j_2$ to the values 0, 2, and 4, with the additional
restriction that $j_1 + j_2 \le 6$, and by limiting $v_1$ and
$v_2$ to the values 0 and 1.  We also assume that the three
vibrational manifolds defined by $v_{\text{t}} = v_1 + v_2 = 0$,
1, or 2 are effectively decoupled from one another, which further
reduces the size of the matrix eigenproblem.  The energies of the
$(v, j)$ rovibrational states of the H$_2$ and D$_2$ monomers and
the dimer reduced masses, which appear in the
close-coupled equations for the radial functions $F_\lambda(R)$,
are computed from the parameters listed in Table~\ref{tab:three}.

Because we consider only even values of $j_1$ and $j_2$ here, the
parity of the angular basis function $I_{J,M,\gamma}$ is
controlled by the dimer orbital angular momentum quantum number
$L$; when $L$ is even, $I_{J,M,\gamma}$ has even parity.  Angular
basis functions with different parities are not coupled together
by Eq.~(\ref{eq:ccdimer}).  In addition, for a dimer of two
identical monomers, the overall spatial wave function (exclusive
of spin) must be either symmetric or antisymmetric under exchange
of the two monomers, and the overall spin wave function must also be
symmetric or antisymmetric under monomer exchange.  The total wave
function, which is the product of the spatial and spin wave functions,
must be symmetric or antisymmetric under monomer exchange for
bosonic and fermionic monomers, respectively.

The {\it para\/}-H$_2$ molecule is a
spin-zero composite boson.  For a dimer of such bosons, no
exchange-antisymmetric spin wave function can be constructed,
and therefore only
states whose spatial wave functions are symmetric under monomer exchange
are physically admissible.  These exchange-symmetric spatial wave
functions are
the only (H$_2$)$_2$ wave functions considered here.  On the other hand,
{\it ortho\/}-D$_2$ molecules may have a total nuclear spin quantum
number of either zero or two, and
it is possible to construct ({\it ortho\/}-D$_2$)$_2$ dimers that
have either an exchange-symmetric or an exchange-antisymmetric
spin wave function.
Consequently the spatial wave function for ({\it
ortho\/}-D$_2$)$_2$ may also be either symmetric or antisymmetric
under monomer exchange, provided that the total (spin times
spatial) ({\it ortho\/}-D$_2$)$_2$ wave function is symmetric
under monomer exchange.\cite{danby}

To check that our matrix-based implementation of the
close-coupled formalism is correct, we have used the BOUND
code\cite{bound} to compute the energies of the (H$_2$)$_2$,
H$_2$--D$_2$, and (D$_2$)$_2$ bound states that correlate with
the monomers' $(v, j) = (0, 0)$ ground rovibrational states,
and compare these energies with those
obtained from our matrix-based code.  [Because the BOUND code
employs the rigid-rotor approximation, for this comparison we
ignore the $j$ dependence of the monomer radial wave functions
$\phi_{v,j}(r)$ that appear in Eq.~(\ref{eq:ccdimer}) and replace
these radial wave functions with those for the monomers' ground
rovibrational states.  This is equivalent to neglecting
centrifugal distortion effects on the monomer radial wave
functions.]  The good agreement between these two calculations
confirms the validity of our matrix-based close-coupled approach.

Some of the dimer states discussed below are long-lived quasibound
states that can decay via rotational predissociation.  The
energies reported for these states are those obtained following
the ``infinite wall'' procedure outlined by Grabenstetter and Le
Roy,\cite{grab} in which the energy of the quasibound state is
monitored as $R_{\text{max}}$ is decreased in $0.1~a_0$ steps.
We estimate that using a finite step size of $0.1~a_0$ in this
procedure introduces an uncertainty in the quasibound state
energies of no more than 0.002~cm$^{-1}$.

\section{Empirical adjustments to the potential energy surface}

In this section, we show that if we make two small empirical
modifications to our ab initio potential energy surface, it gives
rotational and rovibrational transition energies for (H$_2$)$_2$,
H$_2$--D$_2$, and (D$_2$)$_2$ dimers in good agreement with those
obtained experimentally. The two modifications involve a small
inward shift of the repulsive wall of the potential energy
surface, which we quantify using an adjustable parameter $s$, and
a slight increase in the magnitude of the surface's $A_{2,2,4}$
term, which we quantify using an adjustable parameter $q$.  The
unmodified, purely ab initio potential energy surface is defined
by $(s, q) = (0, 0)$.

We focus first on H$_2$--D$_2$ and (D$_2$)$_2$ dimer states that
correlate with rotationally cold $(j=0)$ monomers as $R
\rightarrow \infty$. Because the wave functions of these states
are overwhelmingly dominated by angular basis functions $I_{J,M,\gamma}$
with $j_1 = j_2 = 0$ in Eq.~(\ref{eq:ccdimer}), the states'
energies are insensitive to the anisotropic terms ($A_{0,2,2}$,
$A_{2,0,2}$, and $A_{2,2,4}$) of the potential energy surface;
however, the states' energies are very sensitive to the location
of the surface's repulsive wall. We therefore find the optimal
value for $s$ by adjusting $s$ to bring the computed energies for
transitions involving these states into good agreement with
experimentally-measured transition energies.

Next we consider IR-active transitions of the H$_2$--D$_2$ dimer
which involve either (1) a pure vibrational transition $v = 1
\leftarrow 0$ in the H$_2$ monomer and a pure rotational
transition $j = 2 \leftarrow 0$ in the D$_2$ monomer, or (2) a
rovibrational transition $(v, j) = (1, 2) \leftarrow (0, 0)$ in
the H$_2$ monomer and no excitation of the D$_2$ monomer.  These
transitions involve final states whose energies are sensitive to
the $A_{0,2,2}$ and $A_{2,0,2}$ anisotropic terms of the dimer
potential energy surface.  We find that, once the repulsive wall
of the potential energy surface has been shifted inward slightly,
the energies computed for these transitions are in good agreement
with experimental measurements.  This suggests that the
$A_{0,2,2}$ and $A_{2,0,2}$ terms of the shifted potential energy
surface are accurate, at least in the range of $R$ values probed
by the H$_2$--D$_2$ dimer wave functions.

Finally, we examine (H$_2$)$_2$ and (D$_2$)$_2$ dimer states
which correlate with $R \rightarrow \infty$ limits involving one
$j=0$ and one $j=2$ molecule.  Some of these states have energies
that are very sensitive to the strength of the $A_{2,2,4}$ term
of the potential energy surface. By examining how the computed
energies for transitions involving these states change with $q$,
we find the value for $q$ that gives the best overall agreement
with experimental measurements.

\subsection{Combination differences from the H$_2$--D$_2$ and
(D$_2$)$_2$ dimer Q$_1(0)$ infrared spectra}

We begin by computing the $J = 2 \leftarrow 0$ spacings for the
H$_2$--D$_2$ and (D$_2$)$_2$ dimers that correlate with
rotationally cold $(j=0)$ monomers; we perform these computations
both for the dimers' $v_{\text{t}} = 0$ ground vibrational
manifolds and for the $v_{\text{t}} = 1$ manifold accessed by IR
excitation of the
H$_2$ monomer in the H$_2$--D$_2$ dimer.  Accurate experimental
values for these $J = 2 \leftarrow 0$ spacings have been obtained
from a combination-differences analysis of high-resolution
H$_2$--D$_2$ and (D$_2$)$_2$ IR absorption spectra.\cite{mck1} 
Because of the large energy mismatch between the $v = 1$ levels
of H$_2$ and D$_2$, in our calculations we assume that the
H$_2$--D$_2$ dimer states correlating with H$_2$ $(v=1)$ + D$_2$
$(v=0)$ are decoupled from those correlating with H$_2$ $(v=0)$ +
D$_2$ $(v=1)$.  (Strictly
speaking, the $v_{\text{t}} = 1$ dimer states accessed in the IR
absorption experiment are quasibound, and can decay through
vibrational predissociation. Our assumption that these states are
decoupled from the $v_{\text{t}} = 0$ states, however, closes off
this decay channel.  Because the lifetimes of the $v_{\text{t}} =
1$ dimer states are known to be extremely long,\cite{mck1} this
should not materially affect our results.)

Table~\ref{tab:four} shows that the computed $J = 2 \leftarrow 0$
spacings are 0.015~cm$^{-1}$ to 0.025~cm$^{-1}$ lower than the
experimental ones.  If the dimers were rigid rotors, the $J = 2 \leftarrow 0$
spacings would be equal to six times the dimers' respective
rotational constants.  Because the dimers undergo large-amplitude
zero-point motion along the $R$ direction, a rigid-rotor model
for the dimers' overall end-over-end rotation is not really
appropriate.  Nonetheless, this simple-minded picture suggests
that the dimer states supported by the computed potential energy
surface have average intermolecular distances that are slightly
too large, by about $0.02~a_0$ for the (D$_2$)$_2$ dimer and
$0.03~a_0$ for the H$_2$--D$_2$ dimer.

As we noted in our discussion of Fig.~\ref{fig:repul}, a small
inward shift of the repulsive wall of our potential energy
surface would bring it into closer agreement with a
surface\cite{sk} that gives accurate second virial coefficients
for low-temperature H$_2$ gas; such a shift would also reduce
slightly the average intermolecular distances of the H$_2$--D$_2$
and (D$_2$)$_2$ dimers, possibly bringing the computed $J = 2
\leftarrow 0$ spacings into better agreement with experiment. (We
note here that the potential energy surface presented in
Ref.~\onlinecite{sk} was itself obtained by a similar inward
shift of the repulsive wall of an ab initio computed potential
energy surface.) We therefore modify our ab initio H$_2$--H$_2$
potential energy surface as follows.  For $R$ values below
$6.5~a_0$, we shift our computed ab initio interaction energies
to new, smaller, $R$ values defined by
   \begin{equation}
   R_{\text{new}} = R_{\text{old}} - s (6.5~a_0 - R_{\text{old}}) \ ,
   \end{equation}
and then construct a smooth $s$-dependent H$_2$--H$_2$ potential
energy surface (as described above in Sec.~\ref{sec:smooth})
using the shifted points.  Because we have not yet changed the
strength of the $A_{2,2,4}$ term, we are at present implicitly
holding $q$ fixed at $q = 0$.

\begin{figure}
\includegraphics{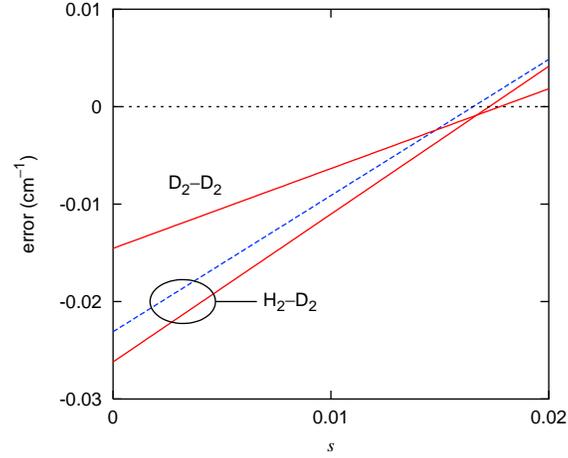}
\caption{(Color online.)
Dependence on $s$ of the errors (computed minus experiment) in the $J = 2 \leftarrow 0$ spacings
of the H$_2$--D$_2$ and (D$_2$)$_2$ dimers.  Solid
lines are for the vibrationally cold ($v_{\text{t}} = 0$) dimers;
the dashed line is for the H$_2$~$(v=1)$ + D$_2$~$(v=0)$ dimer.
The parameter $q$ is held fixed at $q=0$.}
\label{fig:sfig}
\end{figure}

Figure~\ref{fig:sfig} shows how the errors in the $J = 2
\leftarrow 0$ spacings computed for the H$_2$--D$_2$ and
(D$_2$)$_2$ dimers change as $s$ increases from $s = 0$ to $s =
0.025$.  Choosing $s = 0.0175$ brings all three of these computed
spacings into in excellent agreement with experiment. Fixing $s$
at this value amounts to an inward shift of the crossing point
$R_0$, where the vibrationally-averaged H$_2$--H$_2$ isotropic
coefficient $A_{0,0,0}(R) = 0$, from $R_0 = 5.775~a_0$ for the
original ab initio potential energy surface (with $s=0$) to $R_0
= 5.762~a_0$ for the empirically modified surface.  The
corresponding shift for the H$_2$--D$_2$ dimer is from $R_0 =
5.773~a_0$ to $R_0 = 5.760~a_0$; for the (D$_2$)$_2$ dimer, the shift
is from $R_0 = 5.769~a_0$ to $R_0 = 5.756~a_0$.

\subsection{Q$_1(0)$ + S$_0(0)$ and S$_1(0)$ infrared spectra of
the H$_2$--D$_2$ dimer}

\label{sec:h2d2}

Next we consider transitions of the H$_2$--D$_2$ dimer in which
either (1) the H$_2$ monomer undergoes a pure $v = 1 \leftarrow
0$ vibrational transition and the D$_2$ monomer simultaneously
makes a pure $j = 2 \leftarrow 0$ rotational transition, or (2)
the H$_2$ monomer undergoes the rovibrational transition $(v, j)
= (1, 2) \leftarrow (0, 0)$ while the D$_2$ monomer remains in
its rovibrational ground state.  The former transitions belong to
the dimer's Q$_1(0)$~[H$_2$] + S$_0(0)$~[D$_2$] band, and the
latter transitions to the dimer's S$_1(0)$~[H$_2$] band; for
brevity, in this subsection we henceforth drop the molecular
labels in square brackets and simply refer either Q$_1(0)$ +
S$_0(0)$ or S$_1(0)$ transitions.

Because of the large energy mismatch between the $j = 2$ states
of the H$_2$ and D$_2$ molecules, the upper states involved in
these transitions are ones in which the $j = 2$ excitation
remains localized on one of the monomers, and thus have energies
that are insensitive to the $A_{2,2,4}$ term of the potential
energy surface. The computed energies for these transitions
consequently provide insight into the quality of the surface's
$A_{0,0,0}$, $A_{0,2,2}$, and $A_{2,0,2}$ terms.

Four relatively sharp Q$_1(0)$ + S$_0(0)$ transitions and two
relatively sharp $S_1(0)$ transitions have been observed in the
IR absorption spectrum of the H$_2$--D$_2$ dimer.\cite{mck1}  As
Table~\ref{tab:five} shows, the potential energy surface with $s
= 0.0175$ gives transition energies for these six transitions in
very good agreement with experiment; this suggests that the
surface's $A_{0,2,2}$ and $A_{2,0,2}$ terms are fairly accurate,
at least over the range of $R$ values for which the H$_2$--D$_2$
dimer has substantial probability density.

\subsection{S$_0(0)$ infrared spectra of the (H$_2$)$_2$ and
(D$_2$)$_2$ dimers}

Finally we consider IR-active transitions of the (H$_2$)$_2$ and
(D$_2$)$_2$ dimers that correlate with the S$_0(0)$ $j = 2
\leftarrow 0$ pure rotational transitions of the H$_2$ and D$_2$
monomers.  The upper states involved in these transitions are
ones in which the $j = 2$ excitation is shared by the two
monomers; the energies of these states are therefore sensitive to
the $A_{2,2,4}$ term of the dimer potential energy surface, which
couples together angular functions in Eq.~(\ref{eq:ccdimer}) with
$(j_1, j_2) = (0, 2)$ and $(j_1, j_2) = (2, 0)$.

Because of the low reduced mass of the (H$_2$)$_2$ dimer and the
restrictions imposed by nuclear spin statistics, there is just
one sharp S$_0(0)$ IR-active transition for this dimer; it is a
$(J, L) = (1, 1) \leftarrow (0, 0)$ transition and appears in the
(H$_2$)$_2$ far-IR absorption spectrum at
355.425~cm$^{-1}$.\cite{mck3}  The transition energy computed for
this absorption feature using the $(s, q) = (0.0175, 0)$
potential energy surface is 355.438~cm$^{-1}$, or 0.013~cm$^{-1}$
too high.

\begin{figure}
\includegraphics{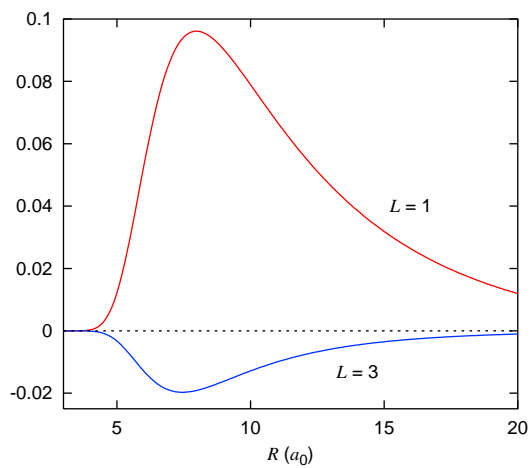}
\caption{(Color online).
The two dominant $F_\lambda(R)$ radial functions for the
$(J, L) = (1, 1)$ upper state accessed in the (H$_2$)$_2$ dimer's
S$_0(0)$ IR transition.  These functions are computed using the
$(s, q) = (0.0175, 0)$ potential energy surface.}
\label{fig:h2sfig}
\end{figure}

The dimer wave function [Eq.~(\ref{eq:ccdimer})] for the upper
state of this transition contains significant contributions from
only four channels: those with $(j_1, j_2, J_{12}, L)$ angular
momentum quantum numbers of $(0, 2, 2, 1)$, $(0, 2, 2, 3)$, $(2,
0, 2, 1)$, and $(2, 0, 2, 3)$. Furthermore, only two of these
channels give independent contributions to the wave function; for
the (H$_2$)$_2$ dimer, exchange symmetry constraints force the
channels with $(j_1, j_2, J_{12}, L) = (a, b, J_{12}, L)$ and
$(j_1, j_2, J_{12}, L) = (b, a, J_{12}, L)$ to, when $L$ is odd,
have radial
functions $F_{\lambda}(R)$ that are equal in magnitude but
opposite in sign.  Figure~\ref{fig:h2sfig} shows
the $F_\lambda(R)$ radial functions for the two independent
channels $(j_1, j_2, J_{12}, L) = (0, 2, 2, 1)$ and $(0, 2, 2,
3)$ that define the upper state of the $(J, L) = (1, 1)
\leftarrow (0, 0)$ transition; about 97\% of the upper state's
probability density is associated with the two $L = 1$ channels.

If we compute the expectation value of the dimer's potential
energy using the upper state wave function,
  \begin{equation}
  \langle V \rangle = \int \int \int \bigl| \Psi({\bf R}, {\bf r}_1, {\bf
  r}_2) \bigr|^2 \, V({\bf R}, {\bf r}_1, {\bf r}_2) \,
  {\rm d}{\bf R} \, {\rm d}{\bf r}_1 \,
  {\rm d}{\bf r}_2 \ ,
  \end{equation}
we find that it includes substantial contributions
from the isotropic $A_{0,0,0}$ term of the potential surface and
the anisotropic $A_{0,2,2}$ and $A_{2,0,2}$ terms, along with a
small contribution from the $A_{2,2,4}$ term; this last
contribution is proportional to the integral
  \begin{equation}
  \int_0^\infty F_{\lambda}(R) F_{\lambda^\prime}(R) {\rm d}R
  \end{equation}
where $F_{\lambda}$ and $F_{\lambda^\prime}$ are
the two radial functions shown in Fig.~\ref{fig:h2sfig}.  As we
explained previously, the lower state for this transition has a
wave function dominated by the $(j_1, j_2) = (0, 0)$ channel, and
its potential energy expectation value is therefore sensitive to
only the isotropic $A_{0,0,0}$ term.

This analysis suggests that a small perturbation of the
$A_{2,2,4}$ term will change the energy of the upper state, but
not that of the lower state, and could thus bring the computed
transition energy for this far-IR absorption feature into better
agreement with experiment. Furthermore, because the transitions
considered in the preceding two subsections involve states whose
energies are insensitive to $A_{2,2,4}$, such a perturbation
would preserve the good agreement with experiment observed for
those transitions.  (Naturally, we could also change the computed
transition energy for this particular far-IR absorption feature
by adjusting the $A_{0,2,2}$ and $A_{2,0,2}$ terms in the
potential energy surface; however, such an adjustment would have
the undesirable side effect of changing the transition energies
computed in the immediately preceding subsection.)  Here we adopt
a very simple adjustment of the $A_{2,2,4}$ term, which helps
compensate for the fact that the $A_{2,2,4}$ coefficients computed
in Sec.~\ref{sec:coeffs} include unwanted contributions from the
electrostatic QH interaction:  we multiply the $A_{2,2,4}$ coefficients computed at each of
the 19 $R$ values by the quantity $(1+q)$, where $q$ is an
adjustable parameter, and then reconstruct the entire potential
energy surface as described in Sec.~\ref{sec:smooth}.

\begin{figure}
\includegraphics{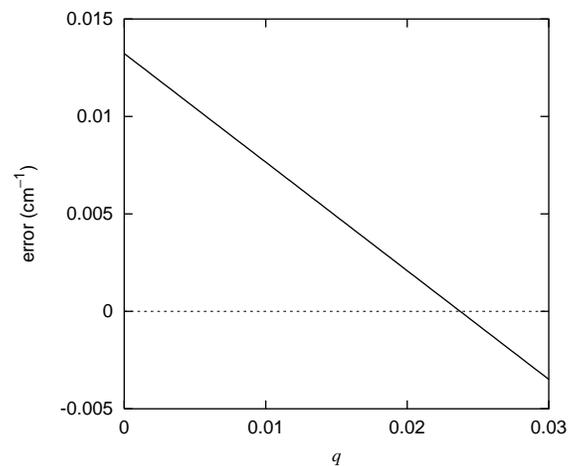}
\caption{Dependence on $q$ of the error (computed minus
experiment) in the transition energy
computed for the (H$_2$)$_2$ dimer's
far-IR S$_0(0)$ absorption feature.  The parameter $s$ is held fixed
at $s = 0.0175$.}
\label{fig:qfig1}
\end{figure}

Figure~\ref{fig:qfig1} shows how $q$ changes the computed
position of the (H$_2$)$_2$ dimer's $(J, L) = (1, 1) \leftarrow
(0, 0)$ S$_0(0)$ far-IR absorption feature. At $q = 0.0235$, the
computed transition energy coincides with the experimental value
of 355.425~cm$^{-1}$.  (However, the uncertainty of $\pm
0.005$~cm$^{-1}$ in this experimental transition energy means
that a wide range of $q$ values would be compatible with the
experimental observations.) This suggests that a simple rescaling
of our $A_{2,2,4}$ coefficients removes much of the QH
interaction's erroneous contribution to these coefficients, even
though the QH interaction has a different power-law dependence on
$R$ than does the quadrupole--quadrupole interaction that
dominates the $A_{2,2,4}$ term.

\begin{figure}
\includegraphics{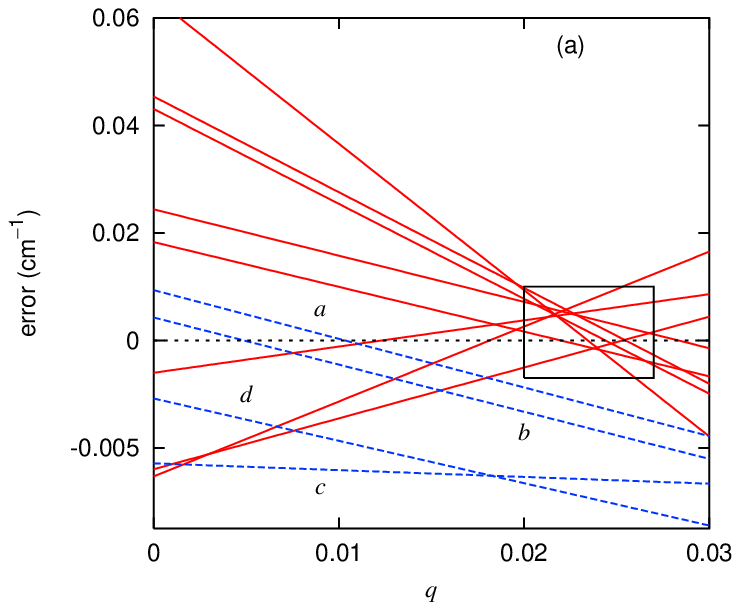}
\includegraphics{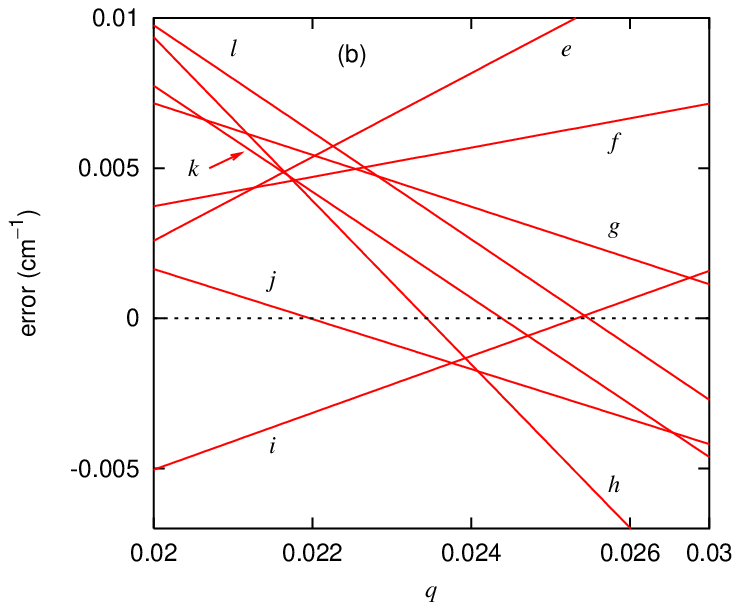}
\caption{(Color online.)
Dependence on $q$ of the errors (computed minus experiment)
in the transition energies
computed for twelve features in the (D$_2$)$_2$ dimer's far-IR
S$_0(0)$ absorption band.  Panel (b) is a magnification of the
small box in panel (a).  The labels affixed to each line
refer to Table~\ref{tab:six}.  The parameter $s$ is held fixed at
$s = 0.0175$.}
\label{fig:qfig2}
\end{figure}

To place tighter constraints on $q$, we turn to the 
(D$_2$)$_2$ dimer, which, because it is both heavier than (H$_2$)$_2$ and
has less severe restrictions arising from nuclear spin
statistics, exhibits many more absorption features in its far-IR
S$_0(0)$ band.\cite{mck3}  Twelve of these features are
relatively sharp, suggesting that they involve bound or
long-lived quasibound states, and also have firmly assigned
initial- and final-state angular momentum quantum numbers.  (We
discuss later a thirteenth sharp transition whose initial- and
final-state assignments are more tentative.)
Figure~\ref{fig:qfig2} and Table~\ref{tab:six} show how the
errors in the energies computed for these twelve transitions
depend on $q$.  For the $q=0$ potential energy surface, the
deviations between computed and measured transition energies
range from $-0.025$~cm$^{-1}$ (transition $e$) to
$+0.064$~cm$^{-1}$ (transition $h$); at $q = 0.0235$, however,
the computed energies for eight of the twelve transitions (those
labeled $e$ through $l$) agree with experiment to within
$\pm0.007$~cm$^{-1}$.  Only one of these eight transitions
has a computed transition energy that differs from the
experimental value by more than $0.005$~cm$^{-1}$, which is the
experimental uncertainty quoted for these transitions in
Ref.~\onlinecite{mck3}.  Furthermore, the value $q = 0.0235$
minimizes the mean absolute deviation between the predicted
and observed transition energies for these eight transitions.

The four transitions labeled $a$ through $d$ in
Table~\ref{tab:six} exhibit poorer agreement with experiment;
furthermore, the errors in the computed energies for these four
transitions at $q = 0.0235$ are equal to or larger in magnitude
than the errors at $q = 0$.  Transition $c$ corresponds to a very
weak far-IR absorption feature; inspection of Fig.~3 in
Ref.~\onlinecite{mck3} shows that its intensity is comparable to
the level of background noise in the (D$_2$)$_2$ absorption
spectrum, and it is possible that the true position of this
feature differs slightly from that reported in
Ref.~\onlinecite{mck3}. Transitions $a$, $b$, and $d$, however,
correspond to relatively strong absorption features; furthermore,
while transition $a$ is a shoulder on the low-energy side of a
very intense feature (transition $l$), transitions $b$ and $d$
are well isolated from other spectral features, and transitions
$a$ and $b$ are linked by the $J = 3 \leftarrow 1$ combination
difference of the dimer's $j_1 = j_2 = 0$ manifold.  It
seems unlikely that the quoted experimental uncertainties for
these three transitions could be badly underestimated. It
therefore appears that our potential energy surface slightly
underpredicts the energies of the $(J, L) = (2, 2)$ and $(3, 3)$
excited-state levels accessed via these three IR transitions.

The experimental (D$_2$)$_2$ S$_0(0)$ IR absorption spectrum
exhibits a thirteenth sharp feature, corresponding to the
transition energy 176.627~cm$^{-1}$, which might be either
the $(J, L) = (3, 1) \leftarrow (2, 2)$ transition or the
$(J, L) = (3, 2) \leftarrow (3, 3)$ transition.\cite{mck3}
Our potential energy surface predicts transition energies of
176.645~cm$^{-1}$ and 176.598~cm$^{-1}$, respectively, for these
transitions.

Finally, we note that the computed transition energies
listed in Table~\ref{tab:five} change by only
0.001 to 0.002~cm$^{-1}$ when the $(s, q) = (0.0175, 0.0235)$
potential energy surface is used.  This validates our decision
to hold $q$ fixed at $q=0$ while we find the optimal value for $s$,
and then hold $s$ fixed at this value while we find the optimal
value for $q$.

\section{Other comparisons with experiment}

In the previous section, we showed that the quality of the four
$A_{l_1,l_2,L}$ terms of our potential energy surface could be
assessed individually by considering transitions between pairs of
states that have energies sensitive to specific subsets of these
terms.  We found that with two small adjustments to the potential
energy surface, we could generate a surface that gives computed
transition energies in fairly good agreement with a number of
high-resolution experimental measurements.

Although some of the transitions considered in the previous
section involve vibrational excitation of the H$_2$ monomer in
the H$_2$--D$_2$ dimer, we have not yet considered vibrationally
excited states of the (H$_2$)$_2$ or (D$_2$)$_2$ dimers.  In
these dimers' $v_{\text{t}} = 1$ vibrationally excited states,
the vibrational excitation is delocalized across the pair of
monomers; transitions to these excited states therefore probe the
simultaneous dependence of the potential energy surface on $r_1$
and $r_2$.

In this section, we show that our modified potential energy
surface predicts energies for these transitions that are in good
agreement with experiment, indicating that the surface accurately
describes the vibrational coupling between the two monomers in
the (H$_2$)$_2$ and (D$_2$)$_2$ dimers. We also consider
IR-active {\it double\/} vibrational transitions of the
(D$_2$)$_2$ dimer, in which each monomer undergoes a $v = 1
\leftarrow 0$ excitation; the good agreement we obtain with
experiment provides further evidence that our modified potential
energy surface has the correct $(r_1, r_2)$ dependence.

\subsection{Q$_1(0)$ spectra of the (H$_2$)$_2$, H$_2$--D$_2$, and
(D$_2$)$_2$ dimers}

Tables~\ref{tab:seven} through \ref{tab:nine} list the energies
of several (H$_2$)$_2$, H$_2$--D$_2$, and (D$_2$)$_2$ bound
states that correlate with $j = 0$ monomer states, computed using
the final $(s, q) = (0.0175, 0.0235)$ potential energy surface. 
Using these bound state energies and the monomer Q$_1(0)$
transition energies from Table~\ref{tab:three}, we can obtain
theoretical positions for the P and R lines in the dimers'
Q$_1(0)$ IR absorption spectra.  In Table~\ref{tab:ten}, we list
the computed positions for the eleven P and R lines that have
been observed experimentally,\cite{mck1} and compare the computed
positions with the observed ones.

The computed transition energies for (H$_2$)$_2$ and H$_2$--D$_2$
are in excellent agreement with experiment, deviating from the
observed energies by amounts smaller than the estimated
experimental uncertainties.  The transition energies for
(D$_2$)$_2$, however, deviate systematically from the
experimental measurements by about $-$0.01~cm$^{-1}$, or about
twice the estimated uncertainty in the measured transition
energies.

\begin{figure}
\includegraphics{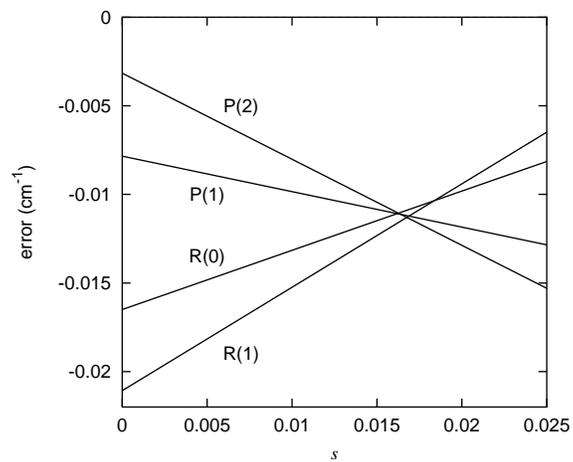}
\caption{Dependence on $s$ of the errors (computed minus experiment)
in the transition
energies computed for four lines in the (D$_2$)$_2$
dimer's Q$_1(0)$ IR absorption band.  The parameter $q$ is held
fixed at $q = 0.0235$.}
\label{fig:qfig3}
\end{figure}

To investigate this discrepancy further, we have computed the
transition energies of the (D$_2$)$_2$ dimer's P(2), P(1), R(0),
and R(1) lines using a set of potential energy surfaces with
different $s$ values, keeping $q$ fixed at $q = 0.0235$. (The P
and R lines involving $J = 3$ states have been omitted from this
analysis simply because computing these states' energies at
several values of $s$ is very
time consuming.)  In Fig.~\ref{fig:qfig3} we show how the
deviations between the computed and observed transition energies
change with $s$.  Only for $s$ values near 0.0175 do the computed
transition energies deviate systematically from experiment; in
addition, the $s = 0.0175$ energies listed in
Table~\ref{tab:nine} give $J = 2 \leftarrow 0$ and $3 \leftarrow
1$ spacings for both the $v_{\text{t}} = 0$ and IR-active
$v_{\text{t}} = 1$ manifolds within 0.002~cm$^{-1}$ of the
experimentally-derived values.\cite{mck1}  These observations
suggest that the systematic deviations observed for (D$_2$)$_2$
in Table~\ref{tab:ten} are not related to a poor choice for
$s$.

These discrepancies could indicate a small error in the isotropic
$A_{0,0,0}$ term's simultaneous dependence on $r_1$ and $r_2$;
the vibrationally excited (D$_2$)$_2$ states involved in the
transitions listed in Table~\ref{tab:ten} are antisymmetric
linear combinations of $(v_1, v_2) = (0, 1)$ and $(1, 0)$ states
and are therefore sensitive to this $(r_1, r_2)$ coupling.
The same coupling term, however, is also active in the IR-active
vibrationally excited (H$_2$)$_2$ state, and in the Raman-active
(H$_2$)$_2$ excited state discussed in the next paragraph, and the
agreement with experiment is excellent for transitions involving
these states of the (H$_2$)$_2$ dimer.  More work is needed to
understand the systematic deviations in the final column of
Table~\ref{tab:nine}.

The (H$_2$)$_2$, H$_2$--D$_2$, and (D$_2$)$_2$ dimers
should all have Raman-active transitions in the vicinity of the
monomers' Q$_1(0)$ Raman transitions; thus far, however, only the (H$_2$)$_2$
dimer's Raman spectrum has been observed
experimentally.\cite{tej}  It consists of a single narrow line
corresponding to a transition energy of 4160.78 $\pm$
0.02~cm$^{-1}$. The theoretical Q$_1(0)$ Raman transition energy
for (H$_2$)$_2$ derived from the first two lines of
Table~\ref{tab:seven} is 4160.764~cm$^{-1}$; the difference
between the computed and experimental Raman transition energies
is only slightly smaller than the estimated experimental
uncertainty.  However, the reported value for the experimental
Raman transition energy is based on a value of 4161.18~cm$^{-1}$
for the Q$_1(0)$ transition of an isolated H$_2$ molecule.  The
difference between this value and the value used here
(4161.169~cm$^{-1}$) accounts for more than half the difference
between the computed and observed dimer Raman transition
energies. If we instead compare the observed and computed
dimerization-induced {\it red shift\/} of the H$_2$ Q$_1(0)$
Raman transition, we find that our computed red shift of
0.405~cm$^{-1}$ is in excellent agreement with the reported
value\cite{tej} of 0.400 $\pm$ 0.02~cm$^{-1}$.

\subsection{S$_1(0)$ and Q$_1(0)$ + S$_1(0)$ infrared spectra of
the (H$_2$)$_2$ and (D$_2$)$_2$ dimers}

We finally use our modified potential energy surface to compute
transition energies for features in the S$_1(0)$ IR absorption
bands of the (H$_2$)$_2$ and (D$_2$)$_2$ dimers, and for features
in the Q$_1(0)$ + S$_1(0)$ IR absorption band of the (D$_2$)$_2$
dimer.  The (D$_2$)$_2$ transitions involve upper states whose
energies are sensitive to the $(r_1, r_2)$ dependence of the
potential energy surface.

The S$_1(0)$ IR absorption spectrum of (H$_2$)$_2$ contains just
one narrow line,\cite{mck1} at $4498.734 \pm 0.004$~cm$^{-1}$.
This feature is associated with a transition from the dimer's
ground state (the first line of Table~\ref{tab:five}) to a state
with $J = L = 1$ that is a linear combination of $(v_1, j_1; v_2,
j_2) = (1, 2; 0, 0)$ and $(0, 0; 1, 2)$; the final state's energy is
listed in the first line of Table~\ref{tab:twelvefive}.
The $(s, q) = (0.0175, 0.0235)$ potential
energy surface gives a computed S$_1(0)$ dimer
transition energy of 4498.729~cm$^{-1}$, in excellent agreement
with the observed value.

The angular basis functions $I_{J,M,\gamma}$ in
Eq.~(\ref{eq:ccdimer}) that correspond to $(J, L, j_1, j_2) = (1,
1, 0, 2)$ and $(1, 1, 2, 0)$, which dominate the final state wave
function for this S$_1(0)$ dimer transition, are not directly coupled together by
any of the four terms $A_{l_1,l_2,L}$ that appear in our
potential energy surface; consequently, this (H$_2$)$_2$
transition, like the H$_2$--D$_2$ transitions considered in
Sec.~\ref{sec:h2d2}, probes primarily the {\it monomer\/}
vibrational dependence of the surface's $A_{0,0,0}$, $A_{0,2,2}$,
and $A_{2,0,2}$ terms. In contrast to the H$_2$--D$_2$
transitions discussed in Sec.~\ref{sec:h2d2}, however, the
(H$_2$)$_2$ S$_1(0)$ transition is sensitive to the $A_{2,2,0}$
and $A_{2,2,2}$ terms of the potential energy surface, which we
have ignored; the fact that we obtain good agreement with
experiment without explicitly including these terms in our
surface is further evidence that these terms are of minor
importance for the dimer bound states considered in this work.

The S$_1(0)$ IR absorption band for (D$_2$)$_2$ is much richer
than that for (H$_2$)$_2$, and is described in
Ref.~\onlinecite{mck1} as ``possibly [the] most informative of
all the hydrogen dimer spectra'' presented there. It contains six
pairs of narrow lines separated by the (D$_2$)$_2$ $v_{\text{t}}
= 0$ ground-state $J = 2 \leftarrow 0$ or $3 \leftarrow 1$
spacings (3.001~cm$^{-1}$ and 4.814~cm$^{-1}$, respectively) and
nine additional narrow lines.

The six pairs of lines are associated with transitions from two
different rotational levels of the $v_{\text{t}} = 0$ ground
state to a common $v_{\text{t}} = 1$ upper state level with a
firm angular momentum quantum number assignment. The upper part
of Table~\ref{tab:eleven} compares the computed and experimental
transition energies for the higher-frequency transition of each
of these pairs.  (No additional information about the quality of
our potential energy surface is carried by the other transition
of each pair.) The agreement between computed and measured
transition energies is quite satisfactory; the largest deviation
is 0.030~cm$^{-1}$, for the $(J, L) = (2, 4) \leftarrow (1, 1)$
transition.

In Ref.~\onlinecite{mck1}, initial- and final-state quantum
number labels were proposed for the nine other narrow lines that
appear in the (D$_2$)$_2$ dimer's S$_1(0)$ IR absorption band;
these assignments were descrbed as ``less certain'' than the
assignments for the pairs of lines linked by ground-state
combination differences.  Eight of these lines are listed in the
lower part of Table~\ref{tab:eleven}, which shows that using
these transition assignments, we again observe very good
agreement between computed and measured transition energies. (The
one line omitted from Table~\ref{tab:eleven} involves a
transition to a $J = 4$ state whose energy we have not attempted
to compute.)  Table~\ref{tab:eleven} thus confirms the transition
assignments proposed in Ref.~\onlinecite{mck1}.

The upper-state wave functions for the transitions listed in
Table~\ref{tab:eleven} are linear combinations of radial
functions in Eq.~(\ref{eq:ccdimer}) with angular momentum quantum
numbers $(J, L, j_1, j_2) = (J, L, 0, 2)$ and $(J, L, 2, 0)$. For
$L \ge 2$, these pairs of radial functions are coupled together
by the $A_{2,2,4}$ term of the potential energy surface. In
addition, some of the upper states accessed via these transitions
are mixtures of S$_1(0)$ states, in which the
rotational and vibrational excitation reside on the same D$_2$
monomer, and Q$_1(0)$ + S$_0(0)$ states, in which one monomer is
vibrationally excited while the other is rotationally excited;
the quantity $F$ listed in Table~\ref{tab:eleven} measures the
degree of mixing in the upper-state wave functions.  The good
agreement between computed and observed transition energies in
Table~\ref{tab:eleven}, especially for transitions to upper
states with $L \ge 2$ or with $F$ values below 0.9, indicates that
the $(r_1, r_2)$ dependence of our potential energy surface, and
of the $A_{2,2,4}$ term in particular, is reasonably accurate.

Further evidence that the $(r_1, r_2)$ dependence of our
potential energy surface is accurate comes from
Table~\ref{tab:twelve}, where we compare the computed and
observed transition energies for several (D$_2$)$_2$ transitions
in the dimer's Q$_1(0)$ + S$_1(0)$ overtone IR absorption band. 
The agreement between computed and observed transition energies
is fairly good, although it appears that the potential energy
surface generally underestimates slightly the energies of the
upper states of these transitions.

\section{Predictions for not-yet-observed transitions}

In this section, we use our final $(s, q) = (0.0175, 0.0235)$
potential energy surface to predict the energies of some
not-yet-observed IR transitions of the H$_2$--D$_2$ dimer
and some not-yet-observed Raman transitions of the (H$_2$)$_2$
and (D$_2$)$_2$ dimers.  Experimental studies designed to
search for these transitions would help test the accuracy of
the potential energy surface presented here.

The region of the H$_2$--D$_2$ dimer's IR absorption spectrum
associated with Q$_1(0)$ excitation of the H$_2$ monomer has
already been studied experimentally, and as Table~\ref{tab:ten}
shows, our potential energy surface gives accurate transition
energies for the four P and R lines in this portion of the dimer's
IR spectrum.  The dimer should have four additional IR-active P and R lines
associated with Q$_1(0)$ excitation of the D$_2$ monomer.  We
have computed the transition energies for these four lines based
on the dimer binding energies listed in Table~\ref{tab:eight};
Table~\ref{tab:fourteen} lists the predicted transition energies
for these four absorption features.

As we noted earlier, the single Raman-active transition in the
(H$_2$)$_2$ dimer's Q$_1(0)$ band was recently observed.  This
dimer should also have Raman-active transitions in the monomer
S$_0(0)$ and S$_1(0)$ bands.  The energies for these
two transitions can be computed from the dimer's ground state
binding energy of 2.894~cm$^{-1}$ and the binding energies of
the $(J, L) = (2, 0)$ states listed in the second line of
Table~\ref{tab:twelvefive}.  We therefore predict that the
(H$_2$)$_2$ dimer should exhibit S$_0(0)$ and S$_1(0)$ Raman
transitions at 354.245~cm$^{-1}$ and 4497.415~cm$^{-1}$,
respectively.  Unfortunately,
these transitions are fairly close to the corresponding
Raman-active transitions of the free H$_2$ monomer, which
are located at transition energies of 354.373~cm$^{-1}$ and
4497.839~cm$^{-1}$, so a high-resolution experiment will likely
be required to observe the dimer transitions.

Because the (D$_2$)$_2$ dimer has four bound states, its Raman
spectrum will be much richer than that of (H$_2$)$_2$.  In
Tables~\ref{tab:fifteen} and \ref{tab:sixteen} we give
predictions for Raman-active transitions of the (D$_2$)$_2$
dimer in the monomer Q$_1(0)$, S$_0(0)$, and S$_1(0)$ bands; these
predictions are based on the energy levels listed in
Tables~\ref{tab:nine} and \ref{tab:thirteen}.  There will
be additional Raman features in the dimer's S$_0(0)$ and S$_1(0)$ bands,
associated with transitions to final dimer states with $J = 3$,
which we have omitted from Table~\ref{tab:sixteen} because we
have not computed the energies of these final dimer states.

\section{Summary and discussion}

We have presented a six-dimensional H$_2$--H$_2$ potential energy
surface that accurately describes several bound (and
long-lived quasibound) states of the (H$_2$)$_2$, (D$_2$)$_2$,
and H$_2$--D$_2$ dimers that correlate with H$_2$ and D$_2$
monomers in their $(v, j) = (0, 0)$, $(0, 2)$, $(1, 0)$, and $(1,
2)$ rovibrational states. The surface is based on a set of ab
initio H$_2$--H$_2$ interaction energies that appear to be nearly
converged with respect to the one-electron and many-electron
basis sets, and which cover fairly densely the region of
configuration space associated with the dimer's van der Waals
well.  The surface incorporates two empirical adjustments: one
softens slightly the surface's repulsive wall at small
H$_2$--H$_2$ distances, and one increases slightly the magnitude
of the surface's $A_{2,2,4}$ term that couples the rotational
degrees of freedom of the two monomers.  The latter adjustment
appears to compensate for the fact that our original $A_{2,2,4}$
coefficients include small contributions from the electrostatic
quadrupole--hexadecapole interaction between the two molecules.
A Fortran subroutine that evaluates the $(s, q) = (0.0175, 0.0235)$
potential energy surface is available through EPAPS.\cite{epaps}

An empirical softening of the ab initio H$_2$--H$_2$ interaction
energy at small intermolecular distances is not unprecedented.
For instance, Schaefer and Kohler\cite{sk}
found that by softening the short-range repulsive wall of an ab
initio H$_2$--H$_2$ potential energy surface, they could bring
properties computed using the surface (in this case, the second
virial coefficient of H$_2$ gas) into better agreement with
experiment.  The surface of Diep and Johnson\cite{jkj1,jkj2} is based
on a series of ab initio calculations that have been extrapolated
to the estimated complete one-electron basis set limit; this
extrapolation technique yields a surface whose repulsive wall is
slightly softer than that of the largest basis-set surface
explicitly calculated, and thus has the same effect as an
empirical softening of the repulsive wall. It would be
interesting to compute the H$_2$--H$_2$ interaction energy using
explicitly-correlated electronic structure methods\cite{r12} to
see whether lingering basis set incompleteness in the present
ab initio calculations is what necessitates the softening of
the short-range repulsive wall of the surface.

We have used our potential energy surface to predict the
energies of 34 not-yet-observed IR and Raman
transitions for (H$_2$)$_2$, (D$_2$)$_2$, and H$_2$--D$_2$
dimers involving even-$j$ states of the H$_2$ and D$_2$ monomers.
Observations of these transitions could help verify the
accuracy of the present potential energy surface, or point
out areas where the surface could be further improved.
Calculations of the energy levels of dimers containing one
$j=1$ molecule, such as {\it ortho\/}-H$_2$--{\it para\/}-H$_2$,
could also be useful in this regard; these calculations
are in progress and will be reported in due course.


\section*{Acknowledgments}

Portions of this work were completed while R.J.H.\ was a
sabbatical visitor at the Laboratory for Physical Chemistry in
the University of Helsinki Department of Chemistry and a
temporary member of the Finnish Center of Excellence in
Computational Molecular Science; the hospitality offered by the
Laboratory's scientific and administrative staff is greatly
appreciated.  A.\ Lignell and M.\ Mets\"al\"a provided valuable
advice regarding the use of University of Helsinki computer
systems, and a grant of computer time from the Center for
Scientific Computing (Espoo, Finland) is gratefully acknowledged.
Several conversations with J.~M.\ Fern\'andez (CSIC/IEM,
Madrid) were also quite helpful in the final stages of this
work.  This manuscript was improved considerably by several
thoughtful suggestions provided by an anonymous referee,
who reviewed it very carefully.

Some of the results reported here were obtained using NWChem
version 4.7, as developed and distributed by Pacific Northwest
National Laboratory (P.~O.\ Box 999, Richland, WA 99352, USA)
and funded by the U.~S.\ Department of Energy.

Financial support was provided by the Air Force Office of
Scientific Research through grant F-49620-01-1-0068, by the
donors of the Petroleum Research Fund, administered by the
American Chemical Society, by the U.S. National Science
Foundation through grant CHE-0414705, by a University of
Tennessee Professional Development Award, and by a grant from the
Suomen Kulttuurirahasto (Finnish Cultural Foundation).


\clearpage


\clearpage

\clearpage

\clearpage

\clearpage

\clearpage

\clearpage


\begin{table} 
\caption{\label{tab:one} Comparison of angular expansion coefficients
$A_{l_1, l_2, L}$ (in cm$^{-1}$) computed using two spherical quadrature
rules at $(R, r_1, r_2) = (4.5~a_0, 1.4~a_0, 1.7~a_0)$.  These coefficients
do not include the full-triples correction.}
\begin{ruledtabular}
\begin{tabular}{ccc}
\noalign{\smallskip}
$(l_1, l_2, L)$ & 18-point & 24-point \\
\hline
(0, 0, 0) & \phantom{$-$}674.559 & \phantom{$-$}674.629 \\
(0, 2, 2) & \phantom{$-$6}32.984 & \phantom{$-$6}33.577 \\
(2, 0, 2) & \phantom{$-$6}20.195 & \phantom{$-$6}20.465 \\
(2, 2, 4) & \phantom{$-$6}19.017 & \phantom{$-$6}19.174 \\
(2, 2, 0) & \phantom{66}$-$0.599 & \phantom{$-$66}1.413 \\
(2, 2, 2) & \phantom{$-$66}0.465 & \phantom{66}$-$0.292 \\
\end{tabular}
\end{ruledtabular}
\end{table}

\clearpage


\begin{table} 
\caption{\label{tab:two} Angular expansion coefficients
$A_{l_1, l_2, L}(R, r_1, r_2)$, in cm$^{-1}$, computed from aug-cc-pVTZ
and aug-cc-pVQZ ab initio energies.  A (3s3p2d) set of bond functions
is used in all calculations.  The coefficients are grouped into pairs
of rows corresponding to fixed $(R, r_1, r_2)$; the upper row in
each pair lists the aug-cc-pVQZ coefficients, while the lower
row in each pair lists the aug-cc-pVTZ coefficients.  These coefficients
do not include the full-triples correction.}
\begin{ruledtabular}
\begin{tabular}{cccccc}
\noalign{\smallskip}
$R$~($a_0$) & $(r_1, r_2)$~($a_0$) & $A_{0,0,0}$ & $A_{2,0,2}$ & $A_{0,2,2}$ & $A_{2,2,4}$ \\
\hline
4.5 & (1.1, 1.1) & \phantom{$-$}378.145 & \phantom{$-$}10.026 & \phantom{$-$}10.026 & \phantom{1}6.360 \\
    &            & \phantom{$-$}382.457 & \phantom{$-$}10.094 & \phantom{$-$}10.094 & \phantom{1}6.320 \\
4.5 & (1.4, 1.4) & \phantom{$-$}570.418 & \phantom{$-$}19.989 & \phantom{$-$}19.989 & 14.277 \\
    &            & \phantom{$-$}576.072 & \phantom{$-$}20.116 & \phantom{$-$}20.116 & 14.197 \\
4.5 & (1.7, 1.7) & \phantom{$-$}776.340 & \phantom{$-$}31.141 & \phantom{$-$}31.141 & 24.607 \\
    &            & \phantom{$-$}783.772 & \phantom{$-$}31.379 & \phantom{$-$}31.379 & 24.478 \\
4.5 & (1.1, 1.7) & \phantom{$-$}573.576 & \phantom{$-$}11.959 & \phantom{$-$}33.199 & 13.032 \\
    &            & \phantom{$-$}579.204 & \phantom{$-$}11.878 & \phantom{$-$}33.480 & 13.044 \\
\hline
5.0 & (1.1, 1.1) & \phantom{$-$}101.523 & \phantom{$-$1}3.023 & \phantom{$-$1}3.023 & \phantom{1}3.607 \\
    &            & \phantom{$-$}103.343 & \phantom{$-$1}3.046 & \phantom{$-$1}3.046 & \phantom{1}3.585 \\
5.0 & (1.4, 1.4) & \phantom{$-$}174.837 & \phantom{$-$1}6.794 & \phantom{$-$1}6.794 & \phantom{1}8.119 \\
    &            & \phantom{$-$}177.554 & \phantom{$-$1}6.879 & \phantom{$-$1}6.879 & \phantom{1}8.070 \\
5.0 & (1.7, 1.7) & \phantom{$-$}262.487 & \phantom{$-$}11.883 & \phantom{$-$}11.883 & 14.347 \\
    &            & \phantom{$-$}266.336 & \phantom{$-$}12.043 & \phantom{$-$}12.043 & 14.254 \\
5.0 & (1.1, 1.7) & \phantom{$-$}178.370 & \phantom{$-$1}3.853 & \phantom{$-$}12.040 & \phantom{1}7.370 \\
    &            & \phantom{$-$}181.142 & \phantom{$-$1}3.883 & \phantom{$-$}12.179 & \phantom{1}7.325 \\
\hline
6.5 & (1.4, 1.4) & \phantom{1}$-$22.530 & \phantom{1}$-$0.543 & \phantom{1}$-$0.543 & \phantom{1}2.070 \\
    &            & \phantom{1}$-$22.239 & \phantom{1}$-$0.528 & \phantom{1}$-$0.528 & \phantom{1}2.054 \\
\end{tabular}
\end{ruledtabular}
\end{table}

\clearpage


\begin{table} 
\caption{\label{tab:three} Monomer spectroscopic constants (in cm$^{-1}$)
and total masses employed in the dimer bound state calculations.}
\begin{ruledtabular}
\begin{tabular}{lcccc}
\noalign{\smallskip}
Species & $\Delta E$ $(v = 1 \leftarrow 0)$ & $B$ $(v=0)$ & $B$ $(v=1)$ & Mass $(m_{\text{e}})$ \\
\hline
H$_2$ & 4161.169 & 59.0622 & 56.1117 & 3674.3 \\
D$_2$ & 2993.614 & 29.8445 & 28.7908 & 7342.9 \\
\end{tabular}
\end{ruledtabular}
\end{table}



\begin{table} 
\caption{\label{tab:four} Observed and computed spacings
(in cm$^{-1}$) between the $J = 0$ and $J = 2$ states
of H$_2$--D$_2$ and D$_2$--D$_2$ dimers that correlate with $j = 0$ states
of the constituent monomers.  These bound state computations employ the unmodified
$(s, q) = (0, 0)$ potential energy surface.}
\begin{ruledtabular}
\begin{tabular}{lcc}
\noalign{\smallskip}
Dimer & Observed & Computed \\
\hline
H$_2$ $(v=0)$ + D$_2$ $(v=0)$ & 3.848 & 3.822 \\
H$_2$ $(v=1)$ + D$_2$ $(v=0)$ & 3.889 & 3.866 \\
D$_2$ $(v=0)$ + D$_2$ $(v=0)$ & 3.001 & 2.986 \\
\end{tabular}
\end{ruledtabular}
\end{table}



\begin{table}
\caption{\label{tab:five} Observed transition energies and deviations
between observed and computed transition energies (computed
minus experiment), in cm$^{-1}$,  for selected IR-active transitions
of the H$_2$--D$_2$ dimer involving $v = 1 \leftarrow 0$ excitation
of the H$_2$ monomer.  The D$_2$ monomer remains in its $v = 0$
vibrational level during the transition.  The initial and
final states are identified by the angular momentum quantum numbers associated
with the dominant term in the dimer wave function [Eq.~(\ref{eq:ccdimer})];
$j_1$ and $j_2$ are the angular momenta of the H$_2$ and D$_2$ molecules,
respectively.  The computed transition energies are obtained using
the $(s, q) = (0.0175, 0)$ potential energy surface.}
\begin{ruledtabular}
\begin{tabular}{ccc}
\noalign{\smallskip}
($J$, $L$, $j_1$, $j_2$)$^\prime$ $\leftarrow$
($J$, $L$, $j_1$, $j_2$)$^{\prime\prime}$ & Observed & Deviation \\
\hline
(1, 1, 0, 2) $\leftarrow$ (2, 2, 0, 0) & 4337.046 & $+$0.003 \\
(2, 1, 0, 2) $\leftarrow$ (2, 2, 0, 0) & 4337.609 & $+$0.002 \\
(0, 2, 0, 2) $\leftarrow$ (1, 1, 0, 0) & 4342.004 & $+$0.013 \\
(1, 2, 0, 2) $\leftarrow$ (1, 1, 0, 0) & 4342.208 & $+$0.007 \\
(1, 1, 2, 0) $\leftarrow$ (2, 2, 0, 0) & 4494.719 & $-$0.007 \\
(2, 1, 2, 0) $\leftarrow$ (2, 2, 0, 0) & 4495.20\phantom{0} & $+$0.009 \\
\end{tabular}
\end{ruledtabular}
\end{table}

\clearpage


\begin{table}
\caption{\label{tab:six} Observed transition energies and deviations
between observed and computed transition energies (computed
minus experiment), in cm$^{-1}$,  for IR-active
(D$_2$)$_2$ transitions in the D$_2$ S$_0(0)$ band.  The computed
transition energies are obtained using either the $(s, q) = (0.0175, 0)$
or the $(s, q) = (0.0175, 0.0235)$ potential energy surface.}
\begin{ruledtabular}
\begin{tabular}{ccccc}
\noalign{\smallskip}
Label &
($J$, $L$)$^\prime$ $\leftarrow$
($J$, $L$)$^{\prime\prime}$ & Observed & Deviation & Deviation \\
 & & & $q = 0$ & $q = 0.0235$ \\
\hline
$a$ & (2, 2) $\leftarrow$ (3, 3) & 175.507 & $+$0.010 & $-$0.012 \\
$b$ & (2, 2) $\leftarrow$ (1, 1) & 180.322 & $+$0.005 & $-$0.016 \\
$c$ & (2, 1) $\leftarrow$ (2, 2) & 177.359 & $-$0.022 & $-$0.026 \\
$d$ & (3, 3) $\leftarrow$ (2, 2) & 181.287 & $-$0.010 & $-$0.029 \\
\noalign{\smallskip}
\hline
\noalign{\smallskip}
$e$ & (1, 2) $\leftarrow$ (1, 1) & 182.328 & $-$0.025 & $+$0.007 \\
$f$ & (1, 3) $\leftarrow$ (0, 0) & 184.536 & $-$0.006 & $+$0.005 \\
$g$ & (2, 0) $\leftarrow$ (3, 3) & 172.776 & $+$0.024 & $+$0.004 \\
$h$ & (0, 2) $\leftarrow$ (1, 1) & 177.996 & $+$0.064 & $+$0.000 \\
$i$ & (2, 3) $\leftarrow$ (2, 2) & 182.797 & $-$0.024 & $-$0.002 \\
$j$ & (2, 0) $\leftarrow$ (1, 1) & 177.592 & $+$0.018 & $-$0.001 \\
$k$ & (1, 1) $\leftarrow$ (0, 0) & 178.747 & $+$0.043 & $+$0.002 \\
$l$ & (1, 1) $\leftarrow$ (2, 2) & 175.744 & $+$0.045 & $+$0.004 \\
\end{tabular}
\end{ruledtabular}
\end{table}



\begin{table}
\caption{\label{tab:seven} Binding energies (in cm$^{-1}$) of selected (H$_2$)$_2$
dimer states that correlate with $j=0$ states of the constituent monomers.
These bound state computations employ the $(s, q) = (0.0175, 0.0235)$
potential energy surface.}
\begin{ruledtabular}
\begin{tabular}{ccc}
\noalign{\smallskip}
$v_{\text{t}}$ & $J$ & Binding energy \\
\hline
0 & 0 & 2.894 \\
1 & 0 & 3.299 \\
1 & 1 & 1.554 \\
\end{tabular}
\end{ruledtabular}
\end{table}

\clearpage


\begin{table}
\caption{\label{tab:eight} Binding energies (in cm$^{-1}$) of selected H$_2$--D$_2$
dimer states that correlate with $j=0$ states of the constituent monomers.   These bound state computations employ the
$(s, q) = (0.0175, 0.0235)$ potential energy surface.}
\begin{ruledtabular}
\begin{tabular}{cccc}
\noalign{\smallskip}
$J$ & \multicolumn{3}{c}{Vibrational state} \\
 & \multicolumn{3}{c}{\hrulefill} \\
 & $v_{\text{t}} = 0$ & $v = 1$ (H$_2$) & $v = 1$ (D$_2$) \\
\hline
0 & 4.417 & 4.792 & 4.644 \\
1 & 3.074 & 3.442 & 3.299 \\
2 & 0.568 & 0.902 & 0.775 \\
\end{tabular}
\end{ruledtabular}
\end{table}



\begin{table}
\caption{\label{tab:nine} Binding energies (in cm$^{-1}$) of selected (D$_2$)$_2$
dimer states that correlate with $j=0$ states of the constituent monomers.   The letters S and A indicate that
the state is respectively symmetric or antisymmetric under monomer exchange.
These bound state computations employ the $(s, q) = (0.0175, 0.0235)$
potential energy surface.}
\begin{ruledtabular}
\begin{tabular}{cccc}
\noalign{\smallskip}
$J$ & \multicolumn{3}{c}{Vibrational state} \\
 & \multicolumn{3}{c}{\hrulefill} \\
 & $v_{\text{t}} = 0$ & $v_{\text{t}} = 1$ (S) & $v_{\text{t}} = 1$ (A)  \\
\hline
0 & 6.712 (S) & 7.118 & 6.939 \\
1 & 5.696 (A) & 5.925 & 6.101 \\
2 & 3.711 (S) & 4.111 & 3.941 \\
3 & 0.885 (A) & 1.109 & 1.263 \\
\end{tabular}
\end{ruledtabular}
\end{table}

\clearpage


\begin{table}
\caption{\label{tab:ten} Computed transition energies (in cm$^{-1}$) and deviations
from experiment (computed minus experiment)
of the P and R lines in the Q$_1(0)$ IR bands of the (H$_2$)$_2$, H$_2$--D$_2$, and (D$_2$)$_2$
dimers.  For the H$_2$--D$_2$ dimer, the vibrationally excited state correlates
with $v=1$ H$_2$ + $v=0$ D$_2$; for the (H$_2$)$_2$ and (D$_2$)$_2$ dimers, the
vibrational excitation is delocalized antisymmetrically across the two monomers.
The computed
transition energies are obtained using the $(s, q) = (0.0175, 0.0235)$ potential energy surface.}
\begin{ruledtabular}
\begin{tabular}{ccccccc}
\noalign{\smallskip}
$J^\prime \leftarrow J^{\prime\prime}$ & \multicolumn{2}{c}{(H$_2$)$_2$} & \multicolumn{2}{c}{H$_2$--D$_2$} & \multicolumn{2}{c}{(D$_2$)$_2$} \\
 & \multicolumn{2}{c}{\hrulefill} & \multicolumn{2}{c}{\hrulefill} & \multicolumn{2}{c}{\hrulefill} \\
 & Computed & Deviation & Computed & Deviation & Computed & Deviation \\
\hline
$2 \leftarrow 3$ & --- & ---           & --- & ---           & 2990.558 & $-$0.007 \\
$1 \leftarrow 2$ & --- & ---           & 4158.296 & $-$0.003 & 2991.400 & $-$0.012 \\
$0 \leftarrow 1$ & --- & ---           & 4159.451 & $-$0.001 & 2992.371 & $-$0.011 \\
$1 \leftarrow 0$ & 4162.509 & $-$0.004 & 4162.144 & $-$0.003 & 2994.401 & $-$0.011 \\
$2 \leftarrow 1$ & --- & ---           & 4163.341 & $+$0.000 & 2995.368 & $-$0.011 \\
$3 \leftarrow 2$ & --- & ---           & --- & ---           & 2996.216 & $-$0.013 \\
\end{tabular}
\end{ruledtabular}
\end{table}



\begin{table}
\caption{\label{tab:twelvefive} Energies (in cm$^{-1}$) for
(H$_2$)$_2$ states with $J_{12} = 2$, $v_{\text{t}} = 0$
or 1, and $L \le 1$.  The energies are
obtained using the $(s, q) = (0.0175, 0.0235)$ potential energy
surface, and are given relative to the S$_0(0)$ and S$_1(0)$
H$_2$ monomer energies for $v_{\text{t}} = 0$ and 1, respectively.}
\begin{ruledtabular}
\begin{tabular}{ccc}
\noalign{\smallskip}
($J$, $L$) & Energy & Energy \\
& $v_{\text{t}} = 0$ & $v_{\text{t}} = 1$ \\
\hline
(1, 1) & $-$1.842 & $-$2.004 \\
(2, 0) & $-$3.022 & $-$3.318 \\
(2, 1) & $-$1.191 & $-$1.463 \\
\end{tabular}
\end{ruledtabular}
\end{table}

\clearpage


\begin{table}
\caption{\label{tab:eleven} Observed transition energies and deviations
between observed and computed transition energies (computed
minus experiment), in cm$^{-1}$,  for IR-active
(D$_2$)$_2$ transitions in the D$_2$ S$_1(0)$ band.  The computed
transition energies are obtained using the $(s, q) = (0.0175, 0.0235)$ potential energy surface.
The column labeled $F$ indicates the fraction of upper-state probability
associated with functions in Eq.~(\ref{eq:ccdimer}) in which the
$v = 1$ and $j = 2$ molecular excitations reside on the same monomer.}
\begin{ruledtabular}
\begin{tabular}{cccc}
\noalign{\smallskip}
($J$, $L$)$^\prime$ $\leftarrow$
($J$, $L$)$^{\prime\prime}$ & Observed & Deviation & $F$ \\
\hline
(2, 0) $\leftarrow$ (1, 1) & 3164.705 & $-$0.005 & 0.984 \\
(1, 1) $\leftarrow$ (0, 0) & 3166.195 & $-$0.005 & 0.959 \\
(2, 2) $\leftarrow$ (1, 1) & 3167.890 & $-$0.004 & 0.966 \\
(1, 3) $\leftarrow$ (0, 0) & 3169.919 & $-$0.020 & 0.666 \\
(2, 0) $\leftarrow$ (1, 1) & 3170.429 & $-$0.020 & 0.196 \\
(2, 4) $\leftarrow$ (1, 1) & 3173.702 & $+$0.030 & 0.467 \\
\noalign{\smallskip}
\hline
\noalign{\smallskip}
(3, 2) $\leftarrow$ (3, 3) & 3163.378 & $-$0.009 & 0.993 \\
(3, 1) $\leftarrow$ (2, 2) & 3163.707 & $-$0.003 & 0.989 \\
(2, 1) $\leftarrow$ (2, 2) & 3164.281 & $-$0.011 & 0.992 \\
(0, 2) $\leftarrow$ (1, 1) & 3166.340 & $-$0.004 & 0.871 \\
(1, 2) $\leftarrow$ (1, 1) & 3167.392 & $-$0.030 & 0.942 \\
(2, 3) $\leftarrow$ (2, 2) & 3168.343 & $-$0.028 & 0.860 \\
(3, 3) $\leftarrow$ (2, 2) & 3168.737 & $-$0.004 & 0.891 \\
(2, 1) $\leftarrow$ (2, 2) & 3170.931 & $+$0.003 & 0.298 \\
\end{tabular}
\end{ruledtabular}
\end{table}

\clearpage


\begin{table}
\caption{\label{tab:twelve} Observed transition energies and deviations
between observed and computed transition energies (computed
minus experiment), in cm$^{-1}$,  for IR-active
(D$_2$)$_2$ transitions in the D$_2$ Q$_1(0)$ + S$_1(0)$ band.  The computed
transition energies are obtained using the $(s, q) = (0.0175, 0.0235)$ potential energy surface.}
\begin{ruledtabular}
\begin{tabular}{ccc}
\noalign{\smallskip}
($J$, $L$)$^\prime$ $\leftarrow$
($J$, $L$)$^{\prime\prime}$ & Observed & Deviation \\
\hline
(2, 0) $\leftarrow$ (3, 3) & 6152.870 & $-$0.018 \\
(1, 1) $\leftarrow$ (2, 2) & 6155.672 & $-$0.020 \\
(2, 2) $\leftarrow$ (3, 3) & 6155.672 & $-$0.021 \\
(3, 1) $\leftarrow$ (2, 2) & 6156.767 & $-$0.015 \\
(0, 2) $\leftarrow$ (1, 1) & 6157.770 & $-$0.003 \\
(1, 1) $\leftarrow$ (0, 0) & 6158.669 & $-$0.016 \\
(3, 3) $\leftarrow$ (2, 2) & 6161.471 & $-$0.036 \\
(1, 2) $\leftarrow$ (1, 1) & 6162.773 & $-$0.002 \\
(2, 3) $\leftarrow$ (2, 2) & 6163.261 & $-$0.015 \\
(1, 3) $\leftarrow$ (0, 0) & 6164.872 & $-$0.005 \\
\end{tabular}
\end{ruledtabular}
\end{table}



\begin{table}
\caption{\label{tab:fourteen} Predicted transition energies (in cm$^{-1}$) 
of the P and R lines in the Q$_1(0)$ IR band of the H$_2$--D$_2$ dimer,
for vibrationally excited states correlating with $v=0$ H$_2$ + $v=1$ D$_2$.
The predictions are obtained using the $(s, q) = (0.0175, 0.0235)$ potential energy surface.}
\begin{ruledtabular}
\begin{tabular}{cc}
\noalign{\smallskip}
$J^\prime \leftarrow J^{\prime\prime}$  & Energy  \\
\hline
$1 \leftarrow 2$ & 2990.884 \\
$0 \leftarrow 1$ & 2992.044 \\
$1 \leftarrow 0$ & 2994.732 \\
$2 \leftarrow 1$ & 2995.913 \\
\end{tabular}
\end{ruledtabular}
\end{table}

\clearpage


\begin{table}
\caption{\label{tab:fifteen} Predicted transition energies (in cm$^{-1}$) 
for Raman-active transitions in the Q$_1(0)$ band of the (D$_2$)$_2$ dimer.
The predictions are obtained using the $(s, q) = (0.0175, 0.0235)$ potential energy surface.}
\begin{ruledtabular}
\begin{tabular}{cc}
\noalign{\smallskip}
($J$, $L$)$^\prime$ $\leftarrow$
($J$, $L$)$^{\prime\prime}$ & Energy \\
\hline
(1, 1) $\leftarrow$ (3, 3) & 2988.398 \\
(0, 0) $\leftarrow$ (2, 2) & 2990.207 \\
(0, 0) $\leftarrow$ (0, 0) & 2993.208 \\
(1, 1) $\leftarrow$ (1, 1) & 2993.209 \\
(2, 2) $\leftarrow$ (2, 2) & 2993.214 \\
(3, 3) $\leftarrow$ (3, 3) & 2993.236 \\
(2, 2) $\leftarrow$ (0, 0) & 2996.215 \\
(3, 3) $\leftarrow$ (1, 1) & 2998.047 \\
\end{tabular}
\end{ruledtabular}
\end{table}



\begin{table}
\caption{\label{tab:sixteen} Predicted transition energies (in cm$^{-1}$) 
for Raman-active transitions in the S$_0(0)$ and S$_1(0)$ bands of the
(D$_2$)$_2$ dimer.
The predictions are obtained using the $(s, q) = (0.0175, 0.0235)$ potential energy surface.}
\begin{ruledtabular}
\begin{tabular}{ccc}
\noalign{\smallskip}
($J$, $L$)$^\prime$ $\leftarrow$
($J$, $L$)$^{\prime\prime}$ & S$_0(0)$ energy & S$_1(0)$ energy \\
\hline
(0, 2) $\leftarrow$ (0, 0) & 183.837 & 3167.595 \\
(2, 0) $\leftarrow$ (0, 0) & 178.713 & 3165.725 \\
(2, 2) $\leftarrow$ (0, 0) & 182.694 & 3168.981 \\
(1, 1) $\leftarrow$ (1, 1) & 178.108 & 3165.208 \\
(1, 3) $\leftarrow$ (1, 1) & 184.568 & 3168.916 \\
(0, 2) $\leftarrow$ (2, 2) & 180.836 & 3164.594 \\
(2, 0) $\leftarrow$ (2, 2) & 175.712 & 3162.724 \\
(2, 2) $\leftarrow$ (2, 2) & 179.693 & 3165.980 \\
(1, 1) $\leftarrow$ (3, 3) & 173.298 & 3160.397 \\
(1, 3) $\leftarrow$ (3, 3) & 179.758 & 3164.105 \\
\end{tabular}
\end{ruledtabular}
\end{table}

\clearpage


\begin{table}
\caption{\label{tab:thirteen} Energies (in cm$^{-1}$) for
(D$_2$)$_2$ states with $J_{12} = 2$, $v_{\text{t}} = 0$
or 1, and $J \le 2$.  The energies are
obtained using the $(s, q) = (0.0175, 0.0235)$ potential energy
surface, and are given relative to the S$_0(0)$ and S$_1(0)$
D$_2$ monomer energies for $v_{\text{t}} = 0$ and 1, respectively.}
\begin{ruledtabular}
\begin{tabular}{cccc}
\noalign{\smallskip}
($J$, $L$) & Exchange & Energy & Energy \\
& symmetry & $v_{\text{t}} = 0$ & $v_{\text{t}} = 1$ \\
\hline
(0, 2) & S & $-$1.942 & $-$5.476 \\
(1, 1) & S & $-$7.030 & $-$6.881 \\
(1, 2) & S & $-$5.620 & $-$4.850 \\
(1, 3) & S & $-$1.238 & $-$3.172 \\
(2, 0) & S & $-$7.066 & $-$7.347 \\
(2, 1) & S & $-$5.445 & $-$5.800 \\
(2, 2) & S & $-$3.085 & $-$4.090 \\
(2, 3) & S & $+$0.017 & $-$1.754 \\
\noalign{\smallskip}
\hline
\noalign{\smallskip}
(0, 2) & A & $-$6.767 & $-$5.719 \\
(1, 1) & A & $-$6.654 & $-$6.846 \\
(1, 2) & A & $-$2.427 & $-$4.692 \\
(1, 3) & A & $-$0.194 & $-$3.138 \\
(2, 0) & A & $-$7.172 & $-$7.356 \\
(2, 1) & A & $-$5.752 & $-$5.821 \\
(2, 2) & A & $-$4.457 & $-$4.169 \\
(2, 3) & A & $-$1.813 & $-$1.844 \\
\end{tabular}
\end{ruledtabular}
\end{table}

\clearpage


\end{document}